\documentclass{jfm}

\usepackage{graphicx}
\usepackage{newtxtext}
\usepackage{newtxmath}
\usepackage{natbib}
\usepackage{hyperref}
\hypersetup{
    colorlinks = true,
    urlcolor   = blue,
    citecolor  = black,
}

\newcommand{\RomanNumeralCaps}[1]
\linenumbers

\usepackage[dvipsnames]{xcolor}

\usepackage{caption}
\usepackage{bm}

\usepackage{subcaption}

\usepackage{soul}
\usepackage[normalem]{ulem}
\newcommand\redsout{\bgroup\markoverwith{\textcolor{red}{\rule[0.5ex]{2pt}{0.4pt}}}\ULon}

\def \be {\begin{equation}}
\def \ee {\end{equation}}
\def \bsea{\begin{subeqnarray}} 
\def \esea{\end{subeqnarray}} 
\def \mi {\mathrm{i}}

\title{Internal shear layers in librating spherical shells: the case of attractors}

\author{Jiyang He\aff{1}
  \corresp{\email{jiyanghe123@gmail.com}},
  Benjamin Favier\aff{1},
  Michel Rieutord\aff{2}
 \and St\'ephane Le Diz\`es\aff{1}}

\affiliation{\aff{1}Aix Marseille Univ, CNRS, Centrale Marseille, IRPHE, Marseille, France
\aff{2}IRAP, Universit\'e de Toulouse, CNRS, UPS, CNES, 14 avenue \'Edouard Belin, F-31400 Toulouse, France}

\begin{document}
\maketitle

\begin{abstract}
Following our previous work on periodic ray paths \citep{heInternalShearLayers2022}, we study asymptotically and numerically the structure of internal shear layers for very small Ekman numbers in a three-dimensional (3D) spherical shell and in a two-dimensional (2D) cylindrical annulus when the rays converge towards an attractor. 
We first show that the asymptotic solution obtained by propagating the self-similar solution generated at the critical latitude on the librating inner core describes the main features of the numerical solution.  
The internal shear layer structure and the scaling for its width and velocity amplitude in $E^{1/3}$ and $E^{1/12}$ respectively are recovered. 
The amplitude of the asymptotic solution is shown to decrease to $E^{1/6}$ when it reaches the attractor, as it is also observed numerically. 
However, some discrepancies are observed close to the particular attractors along which the phase of the wave beam remains constant.  
Another asymptotic solution close to those attractors is then constructed using the model of \cite{ogilvieWaveAttractorsAsymptotic2005}. The solution obtained for the velocity has an $O(E^{1/6})$ amplitude, but a different self-similar 
structure than the critical-latitude solution. It also depends on the Ekman pumping at the contact points of the attractor with the boundaries. We demonstrate that it reproduces correctly the numerical solution. 
Surprisingly, the solution close to an attractor with phase shift (that is an attractor that  touches the axis in 3D or in 2D with a symmetric forcing) is found to be much weaker.   
\end{abstract}

%\begin{keywords}
%\end{keywords}

% {\bf MSC Codes }  {\it(Optional)} Please enter your MSC Codes here

\section{Introduction}
\label{sec:intro}
In rotating flows, inertial waves with a frequency smaller than twice the rotation rate propagate at a fixed angle with respect to the rotation axis \citep{greenspan1968theory}.
The frequency and the angle are preserved when inertial waves reflect on a boundary. 
However, an inertial wave beam may contract or expand as it reflects. 
This linear contraction effect is responsible of  inviscid singularities in the inertial wave field \citep{ogilvie2020internal}.

There are two types of inviscid singularities concerned in the present work. One is at the critical latitude of a sphere where the rays are tangent to the boundary and  where  Ekman pumping blows up \citep{robertsSTABILITYMACLAURINSPHEROID1963}. 
This singularity propagates within the fluid along the tangent critical line at the critical latitude \citep{kerswellInternalShearLayers1995}. 
%The singularities are elongated into the bulk.
When regularised by viscosity, it forms concentrated internal shear layers around the critical line. 
The viscous self-similar solution of \citet{mooreStructureFreeVertical1969} and \citet{thomasSimilaritySolutionViscous1972} is expected to 
describe the viscous structure of these thin layers for small Ekman numbers. 
For a librating spheroid, \citet{ledizesInternalShearLayers2017} derived the singularity strength and the amplitude of the self-similar solution by asymptotically matching the shear layer solution with the Ekman layer solution. 
The self-similar solution was found to be in agreement with direct numerical simulation. 
The same self-similar solution (with the singularity strength and the amplitudes derived in an open domain) was also used to describe the solution on a periodic orbit in a spherical shell geometry \cite[][hereafter HFRL22]{heInternalShearLayers2022}. 
In that case, the solution was obtained  by considering its propagation along the periodic orbit for an infinite number of cycles.
It was found to agree very well with the numerical solutions obtained for low Ekman numbers. 
In particular both the internal shear layer structure and its amplitude scaling in $E^{1/12}$ were recovered by the numerical results  using Ekman numbers as low as $10^{-10}$. 

The singularity obtained from the critical latitude on the outer sphere gives rise to different internal shear layers. These layers are weaker, thicker 
and do not possess a self-similar structure % The thickness of these layers  is $O(E^{1/5})$
\citep{kerswellInternalShearLayers1995,linLibrationdrivenInertialWaves2020}. %, for which the self-similar solution is not expected.
\citet{kidaSteadyFlowRapidly2011} obtained their asymptotic structure for a precessing sphere. 
% by considering the regularised Ekman pumping at the critical latitude. 

Besides  libration and precession which drive the flows through viscosity, non-viscous forcing associated with translating or deforming bodies have also been analysed. Many studies have been performed 
in the context of stratified fluids for applications to tidal flows. Analytic results were obtained for the cylinder and the sphere in an unbounded geometry %Unlike the asymptotic matching of two boundary layer solutions in \citet{ledizesInternalShearLayers2017}, 
%In those case, an inviscid solution is generally derived first and then a viscous damping is added to smooth the singularity of it 
\citep{hurleyGenerationInternalWaves1997a,hurleyGenerationInternalWaves1997,voisinLimitStatesInternal2003} and validated 
experimentally in both 2D  \citep{sutherlandInternalWaveExcitation2002,zhangExperimentalStudyInternal2007} and 3D \citep{flynnInternalWaveExcitation2003,voisinInternalWaveGeneration2011,ghaemsaidi3DStereoscopicPIV2013}.
 \citet{hurleyGenerationInternalWaves1997} and \citet{voisinLimitStatesInternal2003} also showed that, in the far-field, the solution takes the self-similar form predicted by \cite{mooreStructureFreeVertical1969}. 
The singularity strength however varies with respect to the nature of the forcing. \citet{machicoaneInfluenceMultipoleOrder2015} discussed this effect for pulsating and oscillating spheres.

The other inviscid singularity is the attractor in a closed container onto which inertial waves tend to focus \citep{maasGeometricFocusingInternal1995}.  
The presence of such singularities is related to the hyperbolic character of the Poincar\'e equation describing the wave structure:  it leads to an ill-posed Cauchy problem
 except for a few geometries such as the cylinder or the sphere \citep{Rieutord2000}.
Attractors also generate intense internal shear layers, as first observed in a trapezoidal tank for a stably stratified fluid \citep{maasObservationInternalWave1997}. 
The asymptotic structure of these layers was analysed in a forced regime in 2D by  \citet{ogilvieWaveAttractorsAsymptotic2005} (hereafter O05). 
Under a few technical hypotheses, he was able to derive the functional equation describing the inviscid streamfunction and to provide the viscous asymptotic 
expression of the streamfunction  close to the attractor. In particular,  O05 showed that, for his quadrilateral geometry possessing a unique attractor, the main contribution to the solution is associated with the logarithmic singularity of the inviscid streamfunction.  We shall use and adapt his results to our geometry. 
%The self-similar solution of Moore \& Saffman function is still employed to describe the asymptotic structure and performs very well as expected. 
His results were confirmed by a numerical study of an inclined rotating square in \citet{jouveDirectNumericalSimulations2014}. 

In a spherical shell, there may exist both critical-latitude and attractor singularities at the same time. 
In HFRL22, we have considered a case where no attractor was present. We have assumed that the fluid was forced by librating the 
inner core at a frequency  such that inertial waves propagated in a direction oriented at   $45^{\textrm{o}}$ with respect to the vertical.  All the ray trajectories were periodic in that case, and the 
(critical) path issued from the critical latitude on the inner core was just a rectangle in the upper left meridional cut of the shell.  
For other frequencies, the rays issued from the critical latitude are expected to perform a more complex pattern and  possibly converge to an attractor \citep{tilgnerDrivenInertialOscillations1999,ogilvieTidalDissipationRotating2004,ogilvieTidalDissipationRotating2009}.  It is this situation we want to address in the present work. 
We consider the same framework as in HFRL22, where local asymptotic solutions propagated in the volume are compared with global numerical results, but for a frequency for which an attractor is now present. 

The paper is organised as follows. 
The framework is introduced in section \S~\ref{sec:framwork}.  In  \S~\ref{sec:configurations}, we describe the 3D and 2D configurations that we have considered, and provide the governing equations. 
In \S~\ref{sec:numerical_methods}, the numerical method used to integrate the equations for each configuration is explained.  
In section \S~\ref{sec:solution_of_propagation_from_critical_latitude}, we first analyse the wave beams emitted from the critical latitude on the inner core. The asymptotic solution built by propagating the self-similar 
solution is compared to the numerical solution. Discrepancies are observed close to the attractors for some of the cases.  
In section \S~\ref{sec:solution_of_attractor}, we then focus on the solution close to the attractors. We construct an asymptotic solution based on the theory of O05
for an attractor without phase shift in \S \ref{sec:asym_attractor}, and provide a numerical validation in \S \ref{sec:results_attractor}. 
A brief conclusion is finally provided in section \S~\ref{sec:conclusion}.

\section{Framework}\label{sec:framwork}

\subsection{Configurations}\label{sec:configurations}

In this paper, we consider the flow of an incompressible fluid of constant kinematic viscosity $\nu^*$ rotating around the vertical axis $\bm{e}_z$ with a uniform rotation rate $\Omega^*$.
We consider two different configurations.
The first one is the axisymmetric flow filling a three-dimensional (3D) spherical shell, as in HFRL22. 
The other configuration is the two-dimensional (2D) flow, but with three velocity components, between two co-axial cylinders whose axis is horizontal, as in \citet{rieutordAnalysisSingularInertial2002} and \citet{rieutordViscousDissipationTidally2010}.
In the following, geometries, governing equations and forcings are described separately for the two configurations. 

\begin{figure}
\centering
\begin{subfigure}{.33\textwidth}
 \centering
  \includegraphics[width=\textwidth]{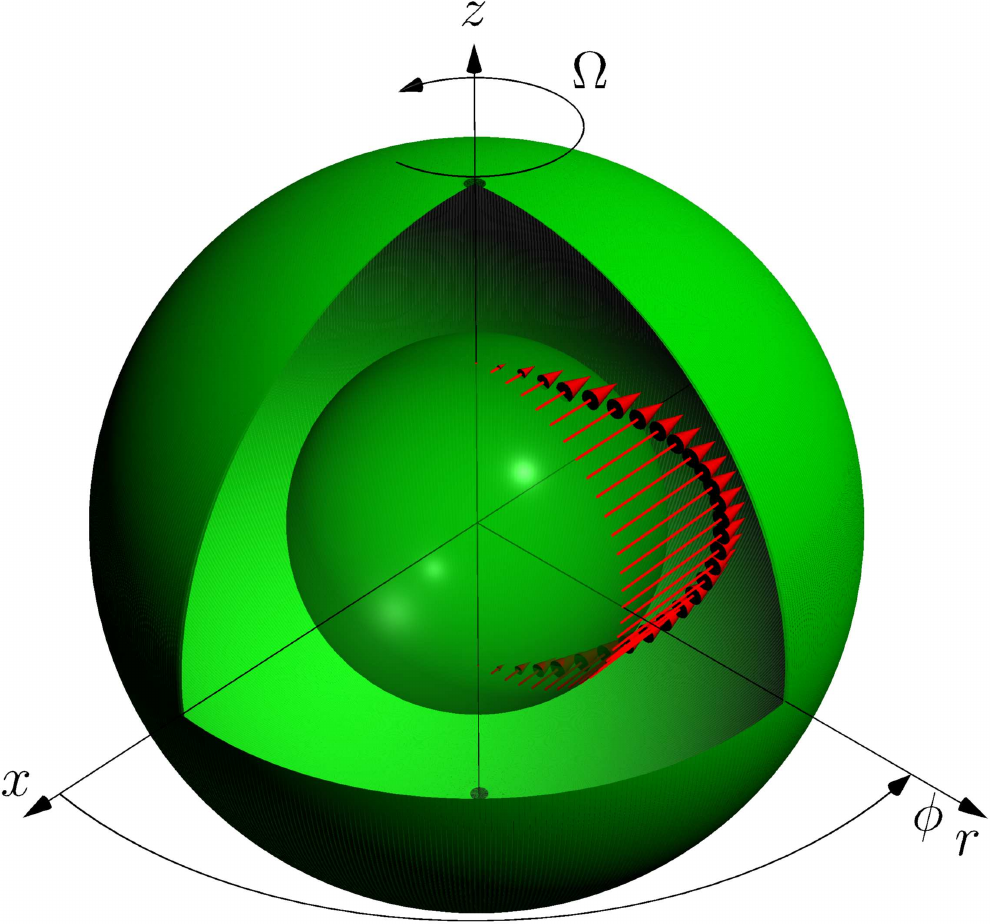}
  \caption{}
  \label{fig:config_3d_libration}
\end{subfigure}%
\begin{subfigure}{.33\textwidth}
\centering
  \includegraphics[width=0.9\textwidth]{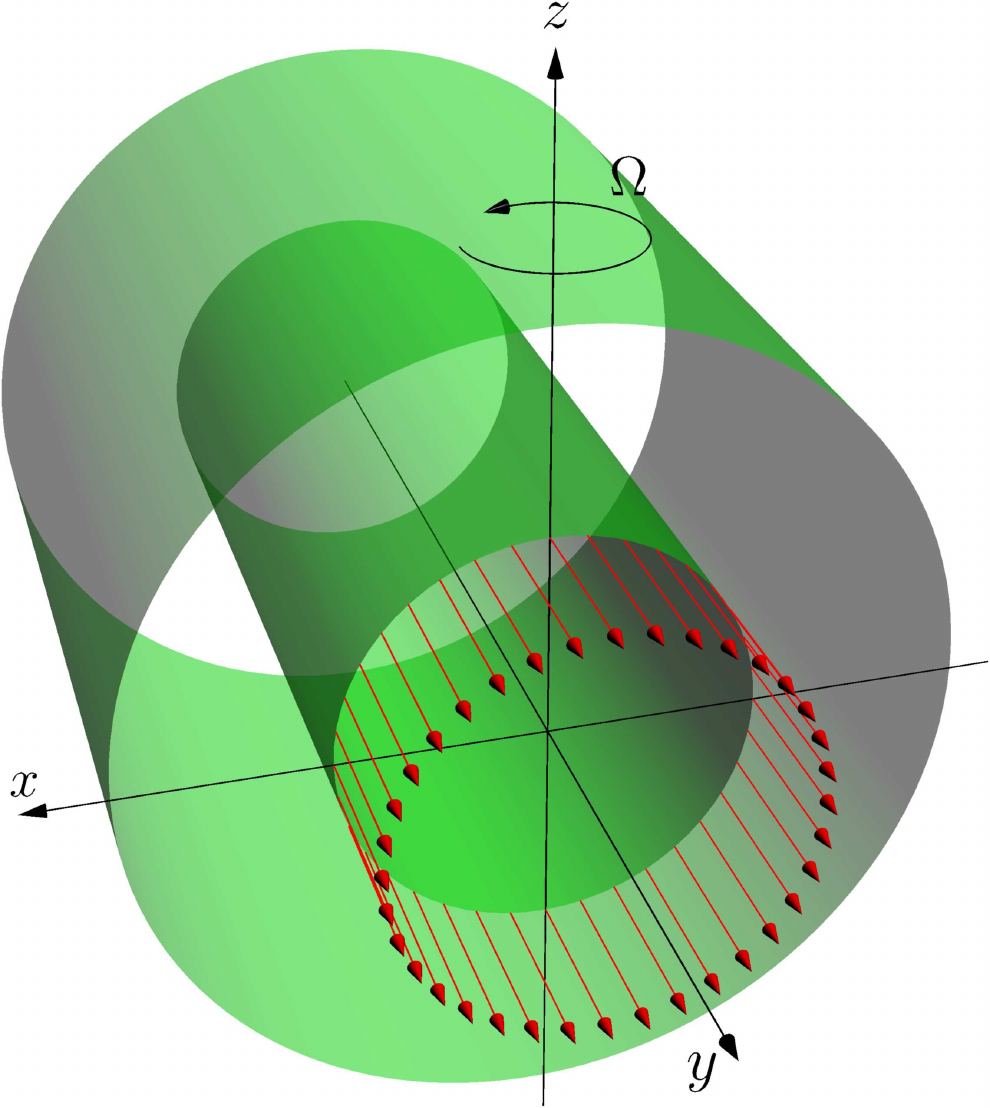}
  \caption{}
  \label{fig:config_2d_symm}
\end{subfigure}
\begin{subfigure}{.33\textwidth}
\centering
  \includegraphics[width=0.9\textwidth]{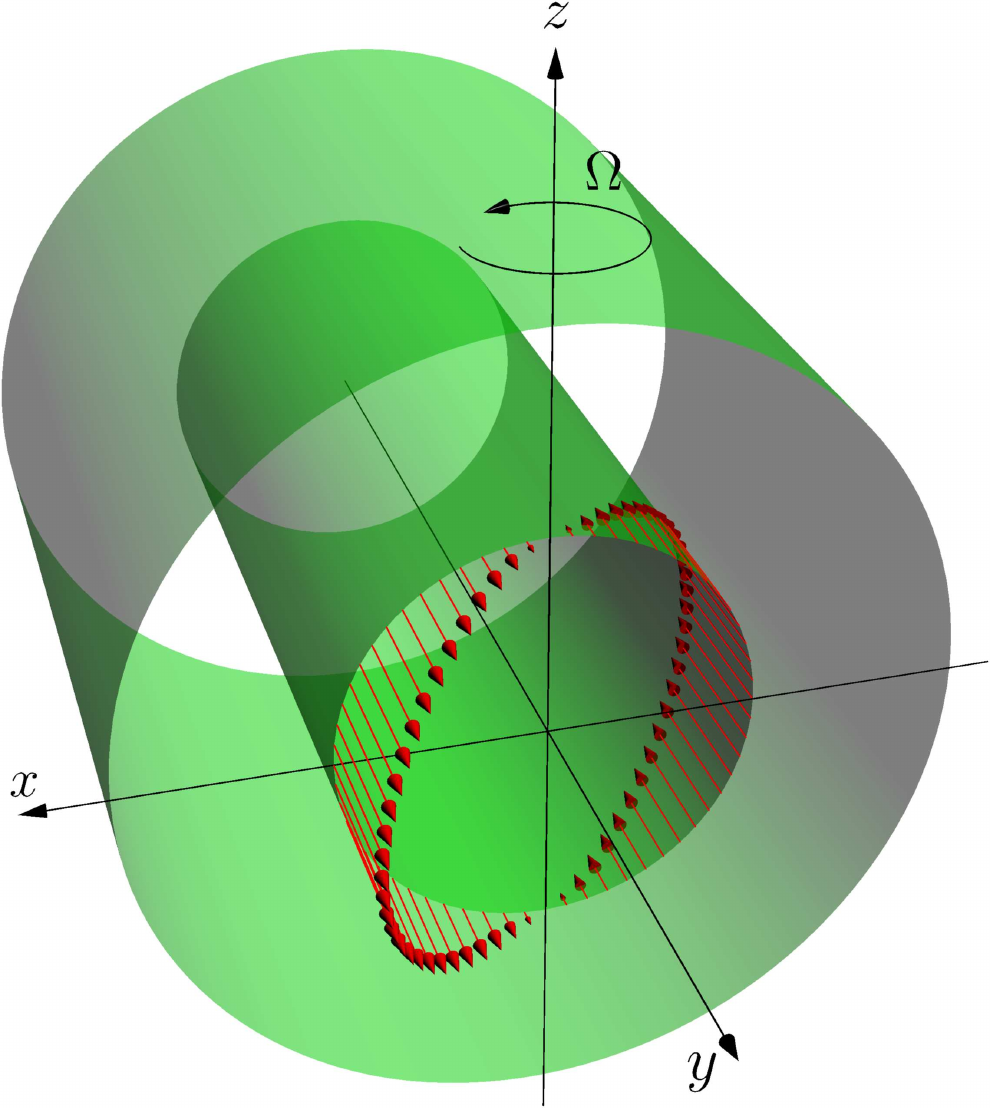}
  \caption{}
  \label{fig:config_2d_anti}
\end{subfigure}
\caption{Configurations: $(a)$ a 3D spherical shell subject to the longitudinal libration on the inner core; $(b)$ a 2D cylindrical annulus subject to the symmetric forcing on the inner core; $(c)$ a 2D cylindrical annulus subject to the antisymmetric forcing on the inner core. The red arrows show the magnitudes and the directions of the forcings at one instant.}
\label{fig:configurations}
\end{figure}

\subsubsection{3D configuration}

The geometry of the 3D spherical shell is shown in figure \ref{fig:config_3d_libration}, whose meridional plane can be found in the figure 2 of HFRL22.
The radii of the outer and inner spheres are $\rho^*$ and $\eta\rho^*$ (with $0<\eta<1$ the aspect ratio), respectively.
Lengths are non-dimensionalised by the outer radius $\rho^*$ such that the inner and outer dimensionless radii are $\eta$ and $1$ respectively.
Time is non-dimensionalised by the angular period $1/\Omega^*$. 
The imposed harmonic forcing is the libration of one of the two boundaries, with the amplitude $\epsilon=\epsilon^*/\Omega^*$ and the frequency $\omega=\omega^*/\Omega^*$. The Ekman number is defined as 
\begin{equation}\label{eq:Ekman_number}
  E=\frac{\nu^*}{\Omega^*\rho^{*2}} \ ,
\end{equation}
with $\nu^*$ being the kinematic viscosity.

As in HFRL22, we care about the linear harmonic response when the Ekman number is extremely small.
We look for solutions that are harmonic in time
\begin{equation}
  \varepsilon(\boldsymbol{v},p)e^{-\mathrm{i}\omega t}+c.c.
\end{equation}
with $c.c.$ denoting complex conjugation.
The velocity $\boldsymbol{v}$ and pressure $p$ satisfy the linearised incompressible Navier-Stokes equations in the rotating frame
\begin{subeqnarray}\label{eq:governing_equations_vector_form}
-\mathrm{i}\omega\boldsymbol{v}+2\boldsymbol{e}_z\times\boldsymbol{v}=-\nabla p+E\nabla^2\boldsymbol{v} \ ,\\[3pt]
\bnabla\bcdot\boldsymbol{v}=0 \ .
\end{subeqnarray}
In terms of the velocity components and pressure, the governing equations in the cylindrical coordinate system $(r,z,\phi)$ become
\begin{subeqnarray}
    -\mi\omega v_r-2v_\phi+\frac{\partial p}{\partial r}-E(\nabla^2-\frac{1}{r^2})v_r=0 \ ,\\
    -\mi\omega v_z+\frac{\partial p}{\partial z}-E\nabla^2v_z=0 \ ,\\
    -\mi\omega v_\phi+2v_r-E(\nabla^2-\frac{1}{r^2})v_\phi=0 \ ,\\
    \frac{\partial v_r}{\partial r}+\frac{v_r}{r}+\frac{\partial v_z}{\partial z}=0 \ ,
\end{subeqnarray}
with the Laplacian operator
\begin{equation}
    \nabla^2=\frac{\partial^2}{\partial r^2}+\frac{1}{r}\frac{\partial}{\partial r}+\frac{\partial^2}{\partial z^2} \ .
\end{equation}

One of the two boundaries is subject to the longitudinal libration as shown by the red arrows in figure \ref{fig:config_3d_libration}, which corresponds to the oscillating solid body rotation of the boundary according to
\begin{equation}\label{eq:libration}
    \boldsymbol{v}(\rho)=r\boldsymbol{e}_\phi \quad \mbox{at} \quad \rho=\eta\quad \mbox{or} \quad 1,
\end{equation}
while the other boundary is subject to the no-slip boundary condition
\begin{equation}
    \boldsymbol{v}(\rho)=\bm{0} \quad \mbox{at} \quad \rho=1\quad \mbox{or} \quad \eta.
\end{equation}
$r$ is the distance to the rotation axis of the cylindrical coordinate system $(r,z,\phi)$, while $\rho$ is distance to the centre in the spherical coordinate system.

\subsubsection{2D configuration}

We also consider a 2D simplification of the 3D axisymmetric configuration discussed above.
The geometry can be viewed as a slender cored torus with the principal radius tending to infinity \citep{rieutordAnalysisSingularInertial2002,rieutordViscousDissipationTidally2010}, which is effectively equivalent to two co-axial cylinders whose principal axis is horizontal, 
as shown in figures \ref{fig:config_2d_symm} and \ref{fig:config_2d_anti}.
The flow between the two cylinders satisfies the similar governing equations as (\ref{eq:governing_equations_vector_form}), while the curvature terms in the differential operators, such as $1/r$, $1/r\partial/\partial r$ and $1/r^2$, are omitted. 
Explicitly, in terms of the velocity components and pressure, the governing equations are
\begin{subeqnarray}
    -\mi\omega v_x-2v_y+\frac{\partial p}{\partial x}-E\nabla^2 v_x=0,\\
    -\mi\omega v_z+\frac{\partial p}{\partial z}-E\nabla^2 v_z=0,\\
    -\mi\omega v_y+2v_x-E\nabla^2 v_y=0,\\
    \frac{\partial v_x}{\partial x}+\frac{\partial v_z}{\partial z}=0,
\end{subeqnarray}
with the Laplacian operator
\begin{equation}
    \nabla^2=\partial^2/\partial x^2+\partial^2/\partial z^2.
\end{equation} 
We use $(x,y,z)$ to denote the Cartesian coordinates, where $Ox$ and $Oz$ are the horizontal and vertical axes respectively and $Oy$ is along the direction perpendicular to the $Oxz$ plane, as shown in figures \ref{fig:config_2d_symm} and \ref{fig:config_2d_anti}.
Note that although we use the same symbol for the Laplacian operators in 2D and 3D, there is no ambiguity since the 2D and 3D operators are independently used in the corresponding dimensions.

The imposed forcing should be viscous and similar to the libration in the 3D configuration. 
The direction of the forcing is thus aligned with that of $\boldsymbol{e}_y$ perpendicular to the $Oxz$ plane.
We consider two options for the amplitude of the forcing.
One option is that the amplitude is a constant, which is
\begin{equation}\label{eq:2d_symmetric_forcing}
    \boldsymbol{v}(\varrho)=\boldsymbol{e}_y \quad \mbox{at} \quad \varrho=\eta\quad \mbox{or} \quad 1,
\end{equation}
where $\varrho=\sqrt{x^2+z^2}$.
The cylinder subject to this forcing is expected to oscillate uniformly along the direction $\boldsymbol{e}_y$, as shown by the red arrows in figure \ref{fig:config_2d_symm}.
The other option is that the amplitude of the forcing depends linearly on the horizontal coordinate $x$, which is
\begin{equation}\label{eq:2d_antisymmetric_forcing}
    \boldsymbol{v}(\varrho)=x\boldsymbol{e}_y \quad \mbox{at} \quad \varrho=\eta\quad \mbox{or} \quad 1.
\end{equation}
The cylinder subject to this forcing oscillates non-uniformly inducing shear at the inner boundary, as shown by the red arrows in figure \ref{fig:config_2d_anti}.
While unrealistic from an experimental point of view, it is a mathematically well-posed boundary condition and provides another symmetry as discussed latter.
While the formula for the 2D antisymmetric forcing (\ref{eq:2d_antisymmetric_forcing}) is similar to the 3D libration case (\ref{eq:libration}), they differ in that the horizontal coordinate $x$ in the 2D configuration can be negative.

Both forcings are symmetric about the horizontal axis $Ox$. However, the former forcing (\ref{eq:2d_symmetric_forcing}) is symmetric about the vertical axis $Oz$, while the latter (\ref{eq:2d_antisymmetric_forcing}) is antisymmetric about $Oz$; see the red arrows in figures \ref{fig:config_2d_symm} and \ref{fig:config_2d_anti} respectively.
These two forcings are thus referred to as symmetric and antisymmetric forcings respectively, according to their symmetries about the $Oz$ axis. They are also imposed on one of the two boundaries, while the other boundary condition is no-slip.

In summary, we consider three different forcings, which are referred to as the 3D libration (\ref{eq:libration}), 2D symmetric (\ref{eq:2d_symmetric_forcing}) and antisymmetric (\ref{eq:2d_antisymmetric_forcing}) forcings. The first one is defined in the 3D spherical shell, while the latter two correspond to the 2D cylindrical annulus.

\subsection{Numerical methods}\label{sec:numerical_methods}
The governing equations (\ref{eq:governing_equations_vector_form}) are solved numerically by spectral methods for both the 3D and 2D configurations. We actually solve the vorticity equation, which is the curl of the momentum equations (\ref{eq:governing_equations_vector_form}$(a)$)
\begin{equation}\label{eq:vorticity_equation}
    -\mathrm{i}\omega\nabla\times\boldsymbol{v}+2\nabla\times (\boldsymbol{e}_z\times\boldsymbol{v})=E\nabla\times(\nabla^2\boldsymbol{v}).
\end{equation}
In the 2D configuration, the curl is only taken in the $Oxz$ plane.
The numerical methods are different for the two configurations. Therefore, they are presented separately.

\subsubsection{3D configuration}\label{sec:numerical_method_3D}
In the 3D configuration, the numerical method is similar to that in our former work (HFRL22). 
The governing equations are solved in the spherical coordinates $(\rho,\theta,\phi)$ with $\rho$ the distance to the centre, $\theta$ the colatitude and $\phi$ the azimuthal angle. 
The velocity is expanded onto the vector spherical harmonics in the angular directions
\begin{equation}
\boldsymbol{v}=\sum_{l=0}^{+\infty}\sum_{m=-l}^{+l}u^l_m(\rho)\boldsymbol{R}_l^m+v^l_m(\rho)\boldsymbol{S}_l^m+w^l_m(\rho)\boldsymbol{T}_l^m,
\end{equation}
with
\begin{equation}
    \boldsymbol{R}_l^m=Y_l^m(\theta,\phi)\boldsymbol{e}_\rho, \quad \boldsymbol{S}_l^m=\nabla Y_l^m, \quad T_l^m=\nabla\times\boldsymbol{R}_l^m.
\end{equation}
The gradients are taken on the unit sphere. 
The vorticity equation (\ref{eq:vorticity_equation}) is projected onto the basis. 
$u^l$ and $w^l$ satisfy a set of ordinary differential equations
\begin{subeqnarray}\label{eq:3d_projection_onto_spherical_harmonics}
    E\Delta_lw^l+\mathrm{i}\omega w^l=-2A_l\rho^{l-1}\frac{\p}{\p \rho}\left(\frac{u^{l-1}}{\rho^{l-2}}\right)-2A_{l+1}\rho^{-l-2}\frac{\p}{\p \rho}\left(\rho^{l+3}u^{l+1}\right), \\[3pt]
    E\Delta_l\Delta_l(\rho u^l)+\mathrm{i}\omega\Delta(\rho u^l)=2B_l\rho^{l-1}\frac{\p}{\p \rho}\left(\frac{w^{l-1}}{\rho^{l-1}}\right)+2B_{l+1}\rho^{-l-2}\frac{\p}{\p \rho}\left(\rho^{l+2}w^{l+1}\right),
\end{subeqnarray}
with
\begin{equation}
    A_l=\frac{1}{l^2\sqrt{4l^2-1}}, \quad B_l=l^2(l^2-1)A_l, \quad \Delta_l=\frac{\mathrm{d}^2}{\mathrm{d}\rho^2}+\frac{2}{\rho}\frac{\mathrm{d}}{\mathrm{d}\rho}-\frac{l(l+1)}{\rho^2},
\end{equation}
\cite[e.g.][]{rieutord91}. Axisymmetry ($m=0$) is employed.
$v^l$ is related to $u^l$ through the continuity equation
\begin{equation}
    v^l=\frac{1}{\rho l(l+1)}\frac{\mathrm{d}\rho^2u^l}{\mathrm{d}\rho}.
\end{equation}
One of the two boundaries is subject to the no-slip boundary condition
\begin{equation}\label{eq:3d_no_slip_boundary_conditions}
    w^l=u^l=\frac{\mathrm{d}u^l}{\mathrm{d}\rho}=0 \quad \mbox{at} \quad \rho=1 \quad \mbox{or} \quad \eta.
\end{equation}
The other boundary is subject to the libration (\ref{eq:libration}), whose projection onto the spherical harmonics yields the inhomogeneous boundary condition
\begin{equation}\label{eq:3d_inhomogeneous_boundary_conditions}
    w^l=2\sqrt{\frac{\pi}{3}}\rho\delta_{1,l}, \quad u^l=\frac{\mathrm{d}u^l}{\mathrm{d}\rho}=0 \quad \mbox{at} \quad \rho=\eta \quad \mbox{or} \quad 1.
\end{equation}
$\delta_{1,l}$ is the Kronecker symbol. Note that the libration is imposed on the spherical harmonic degree $l=1$.

The equations (\ref{eq:3d_projection_onto_spherical_harmonics}-\ref{eq:3d_inhomogeneous_boundary_conditions}) are truncated to the spherical harmonic degree $L$. The derivatives to the radial coordinate $\rho$ are replaced by the Chebyshev differentiation matrices at $N+1$ collocation points of the Gauss-Lobatto grid. Then a block tridiagonal system is obtained as 
\begin{equation}
    \begin{bmatrix}
\boldsymbol{D}_1 & \boldsymbol{C}_1    &         &         &         \\
\boldsymbol{B}_1 & \boldsymbol{D}_2    & \boldsymbol{C}_2     &         &         \\
    & \boldsymbol{\ddots} & \boldsymbol{\ddots}  & \boldsymbol{\ddots}  &         \\
    &        & \boldsymbol{B}_{L-1} & \boldsymbol{D}_{L-1} & \boldsymbol{C}_{L-1} \\
    &        &         & \boldsymbol{B}_L     & \boldsymbol{D}_L      
    \end{bmatrix}
    \begin{bmatrix}
        \boldsymbol{w}^1\\
        \boldsymbol{\rho u}^2\\
        \boldsymbol{\vdots}\\
        \boldsymbol{w}^{L-1}\\
        \boldsymbol{\rho u}^{L}
    \end{bmatrix}=
    \begin{bmatrix}
        \boldsymbol{b}_1\\
        \boldsymbol{b}_2\\
        \boldsymbol{\vdots}\\
        \boldsymbol{b}_{L-1}\\
        \boldsymbol{b}_{L}
    \end{bmatrix}.
\end{equation}
The blocks within the coefficient matrix and the vectors  are $(N+1)\times(N+1)$ and $(N+1)\times 1$, respectively. 
The order of the coefficient matrix is $(N+1)L$ and the number of non-zero elements is $(N+1)^2(3L-2)$. This block tridiagonal system is usually solved by a LU solver \citep{rieutordInertialWavesRotating1997}, by which the coefficient matrix is stored in the banded matrix format and the number of elements in memory is $(N+1)^2(4L-4)-(N+1)(L-2)$. 
% The number of zero elements saved by the band-matrix format is $N(N+1)(L-2)$, which is substantial with high resolution.
On the other hand, the block tridiagonal system can be solved by the block version of the standard tridiagonal algorithm (also called Thomas algorithm), which is the Gaussian elimination on a block tridiagonal system.
This method has been utilised by \citet{ogilvieTidalDissipationRotating2004}.
The algorithm can be found in \citet{engeln-meullgesNumericalAlgorithms1996} (p.121). 
The elimination is advanced forward from the lowest spherical harmonic degree to the highest and the block tridiagonal matrix is reduced to a block upper bidiagonal one, then the solution is obtained by backward substitution. 
During the forward elimination, the updated diagonal block $\boldsymbol{D}_l$ is factorized by the LU solver. A partial pivoting of the block is employed in order to improve the numerical stability.

The three blocks $\boldsymbol{B}_l$, $\boldsymbol{D}_l$ and $\boldsymbol{C}_l$ and the inhomogeneous term $\boldsymbol{b}_l$ at the spherical harmonic degree $l$ are only needed when they take part in the forward elimination. 
Hence, the storage of the whole coefficient matrix is unnecessary.
However, all the updated super diagonal blocks $C_l$ should be reserved in memory for the backward substitution. Their size is $(N+1)^2(L-1)$,
which is almost one third of that of non-zero elements in the original coefficient matrix and one fourth of that in the banded matrix format required by the global LU solver. 
Therefore, the memory usage of the block tridiagonal algorithm is much less than that of the global LU solver, especially when $L$ and $N$ are very large, as required for very low Ekman numbers. 
% In terms of the algorithm complexity, the number of float-point operations required by the block tridiagonal algorithm is also almost one fourth of that required by the banded LU solver.
We develop a code based on the block tridiagonal algorithm using the efficient dynamic programming language Julia \citep{bezanson2017julia}. 
For now, we can reach $E=10^{-11}$ by using $8000$ spherical harmonics and $2500$ Chebyshev polynomials  using double precision floating-point format. The memory footprint is around $750GB$.

\subsubsection{2D configuration}\label{sec:numerical_method_2D}

In the 2D configuration, we take the numerical method similar to that adopted by \citet{rieutordAnalysisSingularInertial2002} and \citet{rieutordViscousDissipationTidally2010}.
The vorticity equation (\ref{eq:vorticity_equation}) is solved in the polar coordinates $(\varrho,\vartheta)$ with $\varrho$ the distance to the centre and $\vartheta$ the angle measured from the horizontal axis $Ox$. 
In terms of the streamfunction $\psi$ and the associated variable $\chi$
\begin{equation}\label{eq:2d_streamfunction-definition_polar}
    v_\varrho=-\frac{1}{\varrho}\frac{\partial\psi}{\partial\vartheta}, \quad v_\vartheta=\frac{\partial\psi}{\partial\varrho}, \quad v_y=\chi,
\end{equation}
the vorticity equation is recast to
\begin{subeqnarray}\label{eq:2d_governing_equations_streamfunction_polar}
    -\mathrm{i}\omega\nabla^2\psi+2(\sin{\vartheta}\frac{\partial\chi}{\partial \varrho}+\frac{\cos{\vartheta}}{\varrho}\frac{\partial\chi}{\partial \vartheta})-E\nabla^4\psi=0,\\[3pt]
    -\mathrm{i}\omega\chi-2(\sin{\vartheta}\frac{\partial\psi}{\partial \varrho}+\frac{\cos{\vartheta}}{\varrho}\frac{\partial\psi}{\partial \vartheta})-E\nabla^2\chi=0,
\end{subeqnarray}
with the operator 
\begin{equation}
    \nabla^2=\frac{\partial^2}{\partial \varrho^2}+\frac{1}{\varrho}\frac{\partial}{\partial \varrho}+\frac{1}{\varrho^2}\frac{\partial^2}{\partial \vartheta^2}.
\end{equation}
The streamfunction $\psi$ and the associated variable $\chi$ are expanded by Fourier series in the angular direction as
\refstepcounter{equation}
$$
  \psi=\sum_{l=-\infty}^{+\infty}\psi_l(\varrho)e^{\mathrm{i}l\vartheta}, \quad
  \chi=-\mathrm{i}\sum_{l=-\infty}^{+\infty}\chi_l(\varrho)e^{\mathrm{i}l\vartheta}.
  \eqno{(\theequation{\mathit{a},\mathit{b}})}\label{eq35}
$$
The projection of the governing equations (\ref{eq:2d_governing_equations_streamfunction_polar}) onto this basis is 
\begin{subeqnarray}
    \mathrm{i}\omega\nabla_l^2\psi_l+(\chi_{l-1}^\prime-\chi_{l+1}^\prime)-\frac{1}{\varrho}\left[(l-1)\chi_{l-1}+(l+1)\chi_{l+1}\right]+E\nabla_l^4\psi_l=0,\\[3pt]
    \mathrm{i}\omega\chi_l+(\psi_{l-1}^\prime-\psi_{l+1}^\prime)-\frac{1}{\varrho}\left[(l-1)\psi_{l-1}+(l+1)\psi_{l+1}\right]+E\nabla_l^2\chi_l=0,
\end{subeqnarray}
with
\begin{equation}
    \nabla_l^2=\frac{\mathrm{d}^2}{\mathrm{d}\varrho^2}+\frac{1}{\varrho}\frac{\mathrm{d}}{\mathrm{d}\varrho}-\frac{l^2}{\varrho^2}.
\end{equation}
The unforced boundary is subject to the no-slip boundary condition
\begin{equation}
    \psi_l=\frac{\mathrm{d}\psi_l}{\mathrm{d}\varrho}=\chi_l=0 \quad \mbox{at} \quad \varrho=1 \quad \mbox{or} \quad \eta.
\end{equation}
The other boundary is subject to the viscous forcings (\ref{eq:2d_symmetric_forcing},\ref{eq:2d_antisymmetric_forcing}).
Both forcings are symmetric about the horizontal axis $Ox$, which leads to
\begin{equation}
    \psi_{-l}=-\psi_l, \quad \chi_{-l}=\chi_l.
\end{equation}
Only the non-negative Fourier components are necessary to be computed.
The symmetric forcing (\ref{eq:2d_symmetric_forcing}) imposes the boundary condition
\begin{equation}
    \psi_l=\frac{\mathrm{d}\psi_l}{\mathrm{d}\varrho}=0, \quad \chi_l=\mathrm{i}\delta_{0,l} \quad \mbox{at} \quad \varrho=\eta \quad \mbox{or} \quad 1.
\end{equation}
Note that the forcing is imposed at $l=0$. Therefore, the following Fourier components are excited
\begin{equation}
    \chi_0,\psi_1,\chi_2,\psi_3,\dots.
\end{equation}
On the other hand, the antisymmetric forcing (\ref{eq:2d_antisymmetric_forcing}) imposes the boundary condition
\begin{equation}
    \psi_l=\frac{\mathrm{d}\psi_l}{\mathrm{d}\varrho}=0, \quad \chi_l=\mathrm{i}\frac{\varrho}{2}\delta_{1,l} \quad \mbox{at} \quad \varrho=\eta \quad \mbox{or} \quad 1.
\end{equation}
Note that the forcing is imposed at $l=1$. Therefore, the following Fourier components are excited
\begin{equation}
    \chi_1,\psi_2,\chi_3,\psi_4,\cdots.
\end{equation}

As in the 3D configuration, the equations are truncated at the Fourier component $L$ and the derivatives to $\varrho$ are replaced by the Chebyshev differentiation matrices with the order $N+1$. 
The resulting block tridiagonal system is solved by the same block tridiagonal algorithm using the efficient dynamic programming language Julia \citep{bezanson2017julia}.

The validation of the two spectral codes used in this paper can be found in the Appendix~\ref{app:verification}.

\section{Wave beams from the critical latitude on the inner core}\label{sec:solution_of_propagation_from_critical_latitude}

The aforementioned forcings are imposed on the inner core. The forcing frequency $\omega$ is chosen in the inertial range  
such that inertial waves propagate at an inclined angle $\theta_c=\arccos{\omega/2}$ relative to the horizontal plane. 
As in HFRL22, two concentrated wave beams are expected to be generated from the 
 critical latitude localised at $(r,z)=(\eta\sqrt{1-\omega^2/4},\eta\omega/2)$ on the inner core. These wave beams 
 travel along the tangential line at the critical latitude in two opposite directions (northward and southward) and reflect
 on the boundaries and form a ray pattern in the spherical shell geometry. 
In general, for a fixed inclined angle $\theta_c$, any ray pattern is composed of the four rays with opposite propagation directions, which are referred to as the northward, outward, southward and inward, as shown in figure \ref{fig:four_ray_directions}.
 In HFRL22, we considered the case where the ray pattern is a simple periodic pattern. Here,  we consider a more general situation 
 where the wave beams converge towards an attractor. 
 Our first objective is to analyse whether an asymptotic solution can be constructed by propagating the 
 self-similar solution describing the concentrated wave beam emitted from the critical latitude, as it was done in HFRL22.
 
In \S \ref{sec:asymptotic_theory}, the asymptotic theory is presented. The properties of the self-similar solution and of the reflection laws are first recalled and 
adapted to the 2D configurations that we also consider before analysing the propagation towards the attractor.   
The asymptotic solution is then compared to numerical results in \S \ref{sec:results}.

\subsection{Asymptotic theory}\label{sec:asymptotic_theory}

\subsubsection{Viscous self-similar solution and scaling}\label{sec:self_similar_solution}

The concentrated ray beams emitted from the critical latitude are associated with an inviscid singularity along the critical ray \citep{LeDizes2023}. It is the viscous smoothing  of this singularity that gives rise  in the limit of small Ekman numbers to a self-similar expression for the dominant wave beam velocity components  \citep{mooreStructureFreeVertical1969}. 

\begin{figure}
    \centering
    \includegraphics[scale=0.5]{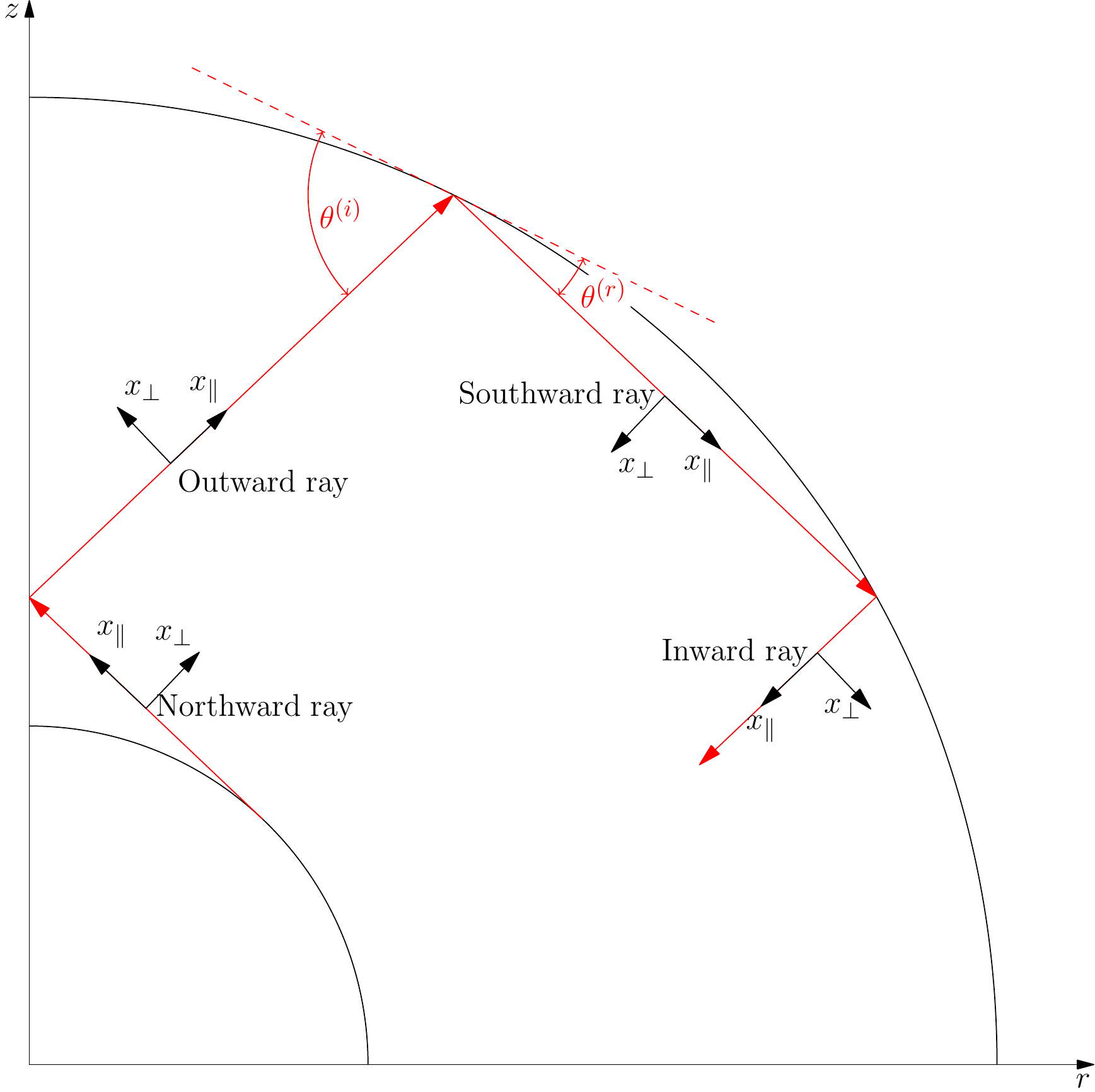}
    \caption{Four propagation directions of the rays in a closed domain. The local vectors attached to each ray are the orientations of the local frames $(x_\parallel,x_\perp)$.}
    \label{fig:four_ray_directions}
\end{figure}

The natural way to describe this self-similar solution is to introduce the local coordinates 
 $(x_\parallel,x_\perp)$ on the critical ray path, with $x_\parallel$ measuring the travelled distance from the source along the critical ray and $x_\perp$ measuring the displacement relative to the critical ray ($x_\perp=0$ is the critical ray equation). %The orientation of $x_\parallel$ is chosen such that the direction of the ray propagation and $x_\parallel=0$ is the location of the source. 
 The orientation of $x_\perp$ is chosen as indicated in figure \ref{fig:four_ray_directions}. It is assumed not to change during the beam propagation. 

The wave beam is centred on the critical ray and has a width of order $E^{1/3}$. In the $(r,z)$ plane, its main velocity component is oriented along ${\boldsymbol{e}_\parallel}$ and can be written at leading order in $E^{1/3}$ in the 3D axisymmetric geometry as  
 \cite[see details in][]{ledizesInternalShearLayers2017}
\begin{equation}\label{eq:self_similar_solution}
    v_\parallel=\frac{1}{\sqrt{r}}C_0H_m(x_\parallel,x_\perp)=\frac{1}{\sqrt{r}}C_0\left(\frac{x_\parallel}{2\sin\theta_c}\right)^{-m/3}h_m(\zeta)
\end{equation}
with the similarity variable
\begin{equation}
    \zeta=x_\perp E^{-1/3}\left(\frac{2\sin\theta_c}{x_\parallel}\right)^{1/3}
\end{equation}
and the special function introduced by \cite{mooreStructureFreeVertical1969} 
\begin{equation}
    h_m(\zeta)=\frac{e^{-\mathrm{i}m\pi/2}}{(m-1)!}\int^{+\infty}_0e^{\mathrm{i}p\zeta-p^3}p^{m-1}dp.
    \label{exp:hm}
\end{equation}
The velocity across the critical rays $v_\perp$ and the pressure $p$ are $O(E^{1/3})$ smaller. 
However, the wave beam  has a velocity component normal to the $(r,z)$ plane of same order which is given by \cite[see][]{ledizesInternalShearLayers2017}
\begin{equation}\label{eq:azimuthal_velocity}
    v_\phi=\pm\mi v_\parallel.
\end{equation}
The sign corresponds to the sign of the projection of the local unit vector $\boldsymbol{e}_\parallel$ onto the global unit vector $\boldsymbol{e}_r$.

The inviscid singularity that gives rise to the self-similar viscous solution is recovered  by taking the limit $\zeta \rightarrow \infty$ in (\ref{eq:self_similar_solution}) 
%The outer limit of the self similar solution (\ref{eq:self_similar_solution}) is \citep{ledizesInternalShearLayers2017}
\begin{equation}\label{eq:outer_limit}
    v_\parallel\rightarrow \frac{1}{\sqrt{r}}C_0x_\perp^{-m}E^{m/3} \quad \mbox{as} \quad \zeta \rightarrow +\infty.
\end{equation}

As we shall see, it is also useful to introduce the streamfunction $\psi$ that can be defined for axisymmetric flows by 
\refstepcounter{equation}
$$
  v_\parallel=\epsilon\frac{1}{r}\frac{\p \psi}{\p x_\perp}, \quad
  v_\perp=-\epsilon\frac{1}{r}\frac{\p \psi}{\p x_\parallel},
  \eqno{(\theequation{\mathit{a},\mathit{b}})}\label{eq:local_velocity_streamfunction_transform}
$$
where $\epsilon=1$ for the rays propagating northward and southward and $\epsilon=-1$ for the rays propagating inward and outward (see figure \ref{fig:four_ray_directions}).
Equation (\ref{eq:local_velocity_streamfunction_transform}a) can be integrated to give at leading order
\begin{equation}\label{eq:self-similar_solution_streamfunction}
    \psi=\epsilon\sqrt{r}\frac{C_0E^{1/3}}{m-1}H_{m-1}(x_\parallel,x_\perp).
\end{equation}
Note that the streamfunction $\psi$ is $E^{1/3}$ smaller than the parallel velocity $v_\parallel$.

The above expressions are valid for 3D axisymmetric geometries. 
For 2D configurations, the term $\sqrt{r}$ is not present in the velocity and streamfunction expressions. 
We get 
\bsea
&&    v_\parallel^{(2D)}=C_0H_m(x_\parallel,x_\perp), \\
 &&    \psi^{(2D)}=\epsilon \frac{C_0E^{1/3}}{m-1}H_{m-1}(x_\parallel,x_\perp).
     \label{eq:2d_self_similar_solution}
\esea
The velocity component $v_y$ perpendicular to the $(x,z)$ plane differs from $v_\parallel^{(2D)}$ by a $\pm \pi/2$  phase factor,  as the relation (\ref{eq:azimuthal_velocity}) between $v_\phi$ and $v_\parallel$ in 3D.

There are free parameters in the self-similar solution (\ref{eq:self_similar_solution}): the singularity strength $m$ and the amplitude $C_0$. 
These parameters depend on the nature of the forcing. 
For a viscous forcing, that is a forcing induced by Ekman pumping, these parameters can be obtained in closed form for the northward and southward beams
generated from the critical latitude \citep{ledizesInternalShearLayers2017,LeDizes2023}. 
For a librating sphere, they are given by \citep{ledizesInternalShearLayers2017} 
\begin{equation}\label{eq:critical_latitude_m}
    m=5/4,
\end{equation}
and
\begin{subeqnarray}\label{eq:critical_latitude_C0}
    C_0=\frac{E^{1/12}}{8(2\sin\theta_c)^{1/2}(-2\kappa_c)^{1/4}}e^{\mathrm{i}\pi/2} \quad \mbox{for the northward beam},\\[3pt]
    C_0=\frac{E^{1/12}}{8(2\sin\theta_c)^{1/2}(-2\kappa_c)^{1/4}}e^{\mathrm{i}3\pi/4} \quad \mbox{for the southward beam},
\end{subeqnarray}
where $\kappa_c$ is the curvature at the critical latitude. 
These expressions can be applied to our geometry for the three forcings (\ref{eq:libration}, \ref{eq:2d_symmetric_forcing} and \ref{eq:2d_antisymmetric_forcing}) imposed on the inner core. 
Considering the different non-dimensionalisation of lengths adopted by \citet{ledizesInternalShearLayers2017} and this work (the radial distance of the critical latitude to the rotation axis vs the outer radius), the absolute value of the complex amplitude $C_0$ should be adapted as indicated in table \ref{tab:absolute_value_of_complex_amplitude} for the three forcings. Note that the curvature at the critical latitude $\kappa_c$ is $-1/\eta$ with  our non-dimensionalisation.  
The factor $\eta\sin{\theta_c}$ is the distance of the critical latitude to the axis $Oz$.

\begin{table}
  \begin{center}
\def~{\hphantom{0}}
  \begin{tabular}{cccc}  
                      & 3D libration  & 2D symmetric forcing   &   2D antisymmetric forcing  \\[5pt]      
       ~$|C_0|$ ~& $\frac{E^{1/12}}{8(2\sin\theta_c)^{1/2}(2/\eta)^{1/4}}(\eta\sin{\theta_c})^{3/2}$  & $\frac{E^{1/12}}{8(2\sin\theta_c)^{1/2}(2/\eta)^{1/4}}$ & $\frac{E^{1/12}}{8(2\sin\theta_c)^{1/2}(2/\eta)^{1/4}}\eta\sin{\theta_c}$ \\ 
  \end{tabular}
  \caption{Absolute value of the complex amplitude $C_0$ for different forcings.}
  \label{tab:absolute_value_of_complex_amplitude}
  \end{center}
\end{table} 

Note that the amplitude $C_0$ of the parallel velocity  scales as $E^{1/12}$.  This scaling has been validated by HFRL22 for Ekman numbers down to $10^{-10}$. The amplitude of the streamfunction is weaker and of order  $E^{5/12}$.

\subsubsection{Reflections on the boundaries and on the axis}\label{sec:reflections}

The reflection of a self-similar wave beam on a boundary has been discussed in \citet{ledizesReflectionOscillatingInternal2020} and HFRL22. 
\citet{ledizesReflectionOscillatingInternal2020} showed that the wave beam keeps its self-similar form when it reflects on a boundary. More precisely if the incident beam  is written as $v_\parallel^{(i)}=C_0^{(i)}H_m(x_\parallel^{(i)},x_\perp^{(i)})$, the reflected beam can also be written as
$v_\parallel^{(r)}=C_0^{(r)}H_m(x_\parallel^{(r)},x_\perp^{(r)})$ with 
\refstepcounter{equation}
$$
  \frac{x_{\parallel _b}^{(r)}}{x_{\parallel b}^{(i)}}=\alpha^3, \quad
  \frac{C_0^{(r)}}{C_0^{(i)}}=\alpha^{m-1},
  \eqno{(\theequation{\mathit{a},\mathit{b}})}\label{eq:reflection_laws}
$$
where the subscript $b$ indicates values taken at the reflection point. The reflection factor $\alpha$ at the reflection point is given 
by
\begin{equation}
\alpha = \frac{\sin \theta^{(r)}}{\sin\theta^{(i)}} 
\end{equation}
where $\theta^{(r)}$ and $\theta^{(i)}$ are the angles of the reflected and incident beams with respect to the boundary (see figure \ref{fig:four_ray_directions}). 
This factor is smaller than 1 (resp. larger than 1) when there is a contraction (resp. expansion) of the beam.  
A reflection on a boundary then just modifies the travelled distance from the source and the amplitude of the beam. In particular it has no effect on its phase. 
 
Note however that this reflection  law implicitly  assumes that the beam is not forced at the boundary where it reflects. 
This in particular implies a simple relation on the streamfunction of the incident and reflected beams at the boundary 
that can be written as
\begin{equation}
\psi^{(r)} (x_{\parallel b}^{(r)},x_{\perp b}^{(r)})  +\psi ^{(i)}(x_{\parallel_b}^{(i)},x_{\perp b}^{(i)}) =0. 
\label{exp:reflection-psi}
\end{equation}
We shall see below that this relation is no longer valid when we get very close to an attractor.

The crossing of the wave beam with the rotation axis is of different nature. In the 3D axisymmetric geometry, the self-similar solution diverges on
the axis, but it can nevertheless be continued as if there was a reflection. The relation between the incident and reflected beams is obtained by a condition of
matching with the solution obtained close to the axis  \cite[see][]{ledizesInternalShearLayers2017}. 
We obtain in that case a phase shift of $\pi/2$ between the reflected and incident beams: 
\refstepcounter{equation}
$$
 x_{\parallel _b}^{(r)} = x_{\parallel b}^{(i)}~, \quad     C_0^{(r)}=e^{\mathrm{i}\varphi}C_0^{(i)} ~ ,
  \eqno{(\theequation{\mathit{a},\mathit{b}})}
 \label{eq:reflection_on_axes}
$$
with $\varphi=\pi/2$.

In the 2D configurations, the condition of reflection to apply on the axis $Oz$ is directly related to the property of symmetry of the forcing. 
On the axis $Oz$, the projections of propagation directions of the incident and reflected rays onto the global unit vector $\boldsymbol{e}_x$ are of opposite sign. According to the formula (\ref{eq:azimuthal_velocity}), we then have the following relations 
\refstepcounter{equation}
$$
  v_y^{(r)}=\pm\mi v_\parallel^{(r)}, \quad
  v_y^{(i)}=\mp\mi v_\parallel^{(i)}.
  \eqno{(\theequation{\mathit{a},\mathit{b}})}
$$
For the 2D symmetric forcing (\ref{eq:2d_symmetric_forcing}) where $v_y$ is forced in a symmetric way about the axis $Oz$, we have $v_y^{(r)}=v_y^{(i)}$. Therefore, the parallel velocities are of opposite sign, which means  
\begin{equation}
    \varphi=\pi ,
\end{equation}
in (\ref{eq:reflection_on_axes}).
For the 2D antisymmetric forcing (\ref{eq:2d_antisymmetric_forcing}) with $v_y^{(r)}=-v_y^{(i)}$, the parallel velocity is unchanged
which means that 
\begin{equation}\label{eq:phase_shift_2D_antisymmetric}
    \varphi= 0.
\end{equation}

In order to consider a quarter of the domain in the $(r,z)$ or $(x,z)$ plane, the horizontal axis $Or$ (or $Ox$) has also to be considered as a place of reflection. Applying the same approach, we can easily show that no phase shift is created
between reflected and incident beams on this axis for all the three forcings. 

\subsubsection{Propagation of critical-latitude beams}\label{sec:propagation}

\begin{figure}
    \centering
    \includegraphics[width=\textwidth]{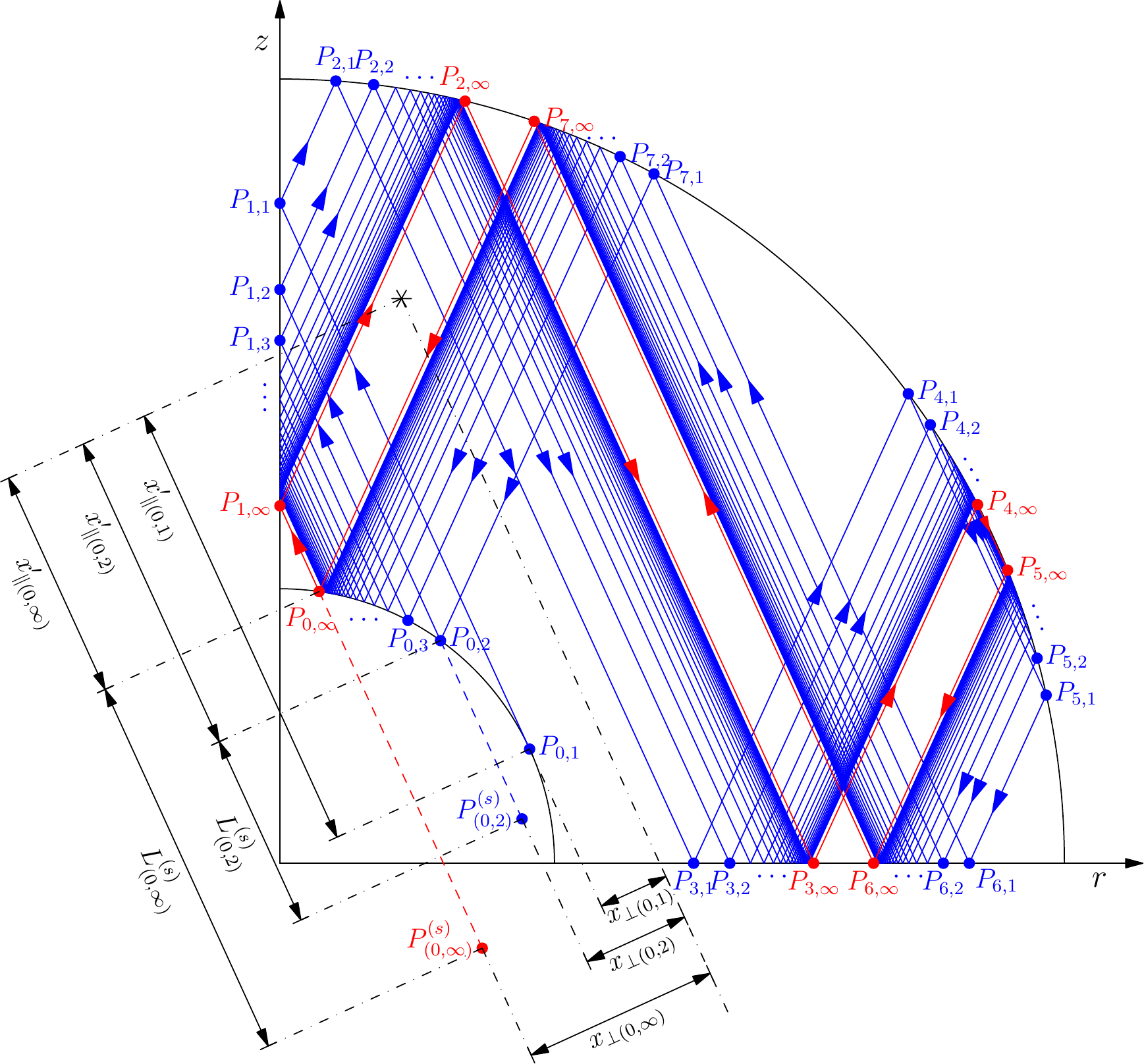}
    \caption{Schematic of propagation of a critical ray towards an attractor for $\eta=0.35, \omega=0.824$}.
    \label{fig:schmatic_of_propagation}
\end{figure}

Having provided the structure of the wave beam and how it reflects on the boundaries and the axis, we are now in a position to 
analyse its propagation in a closed geometry.  
As explained above, we consider a frequency such that the rays emitted from the critical latitude on the inner core
end up on an attractor. Our objective is to obtain the property of the self-similar beam centred on the critical ray 
as it moves towards the attractor. 
An example of critical ray is shown in figure \ref{fig:schmatic_of_propagation} where the ray (blue lines) propagates northward from the critical latitude and spirals into one side of the attractor (red lines). In the following, we use this figure for explanation purposes but 
the methodology is applicable for any type of wave patterns. 
 
The reflection positions on the axes and the boundaries during every loop are indicated as $P_{j,n}$, where  $j$ denotes the reflection position  and ranges from $0$ to $J-1$ ($J=8$ in figure \ref{fig:schmatic_of_propagation}). The index $n$ denotes the number of the cycle and ranges from $1$ to $\infty$. 
For example, the reflection points on the rotation axis are $P_{1,1}$, $P_{1,2}$, $...$ and $P_{1,\infty}$. 
To simplify the formula we assume that the initial point of a cycle is the position $J$ of the former cycle, that is $P_{0,n+1}= P_{J,n}$. 
The critical latitude corresponds to $P_{0,1}$.
The critical ray follows the following path during propagation
\begin{equation}\label{eq:ray_path}
    P_{0,1}P_{1,1}\cdots P_{J-1,1} \rightarrow P_{0,2}P_{1,2}\cdots P_{J-1,2} \rightarrow  \cdots \rightarrow
    P_{0,\infty}P_{1,\infty}\cdots P_{J-1,\infty} .
\end{equation}
The critical ray ends up on the attractor denoted by $P_{0,\infty}P_{1,\infty}\cdots P_{J-1,\infty}$ after an infinite number of cycles. 

The solution obtained by propagating the self-similar beam along the critical ray is expected to be composed of as many contributions as the number of segments
between two reflection points. 
We use the subscript $(j,n)$ to denote the parameters associated with  the segment $P_{j,n}P_{j+1,n}$ (with $j$ between 0 and $J-1$). 
Finding the parameters characterising this contribution requires tracking the variation of the travelled distance and of the amplitude 
during all the previous reflections. 
For this purpose, it is useful to write the travelled distance $x_{\parallel j,n}$ as 
\be
x_{\parallel (j,n)} = L_{j,n}^{(s)} + x_{\parallel (j,n)}' 
\ee
where $x_{\parallel (j,n)}'$ is the distance from $P_{j,n}$ and $L_{j,n}^{(s)}$ is the distance of the source $P_{j,n}^{(s)}$ from $P_{j,n}$. 
The condition of reflection (\ref{eq:reflection_laws}) applied in $P_{j+1,n}$ implies that
\be 
L_{j+1,n}^{(s)} = (L_{j,n}^{(s)} + L_{j,n})\alpha_{j+1,n}^3 ~ ,
\label{exp:Ljn}
\ee
where  $L_{j,n}$ is the length of the segment $(j,n)$ and $\alpha_{j+1,n}$ is the reflection factor at $P_{j+1,n}$.
Concerning the amplitude $C_{j,n}$ of the self-similar solution, we obtain from (\ref{eq:reflection_laws}b)
with  (\ref{eq:critical_latitude_m})
\be
C_{j+1,n} = C_{j,n} \alpha_{j+1,n}^{1/4} e^{\mi \varphi_{j+1}},
\ee
where $\varphi_j $ is the phase shift obtained at the reflection at $P_{j,n}$. For the critical ray shown in figure \ref{fig:schmatic_of_propagation}, this phase shift is null except for $j=1$
(because the reflection is on the axis), for which it can be either $\pi/2$ (3D case), $\pi$ (2D symmetric case) or
0 (2D antisymmetric case). 

In the following, we shall consider the solution in a section perpendicular to the segments $(0,n)$. 
It is therefore useful to consider the evolution of the beam after each cycle for this particular segment as a function of $n$. 
Using (\ref{exp:Ljn}), we can write 
\be
L_{0,n+1}^{(s)} \equiv L_{J,n}^{(s)} = (L_{0,n}^{(s)} + \Lambda _{n} )\alpha_n^3,
\label{exp:L0(s)}
\ee
with 
\begin{equation}\label{eq:length}
\Lambda_n=L_{0,n}+\frac{L_{1,n}}{\alpha_{1,n}^3}+\frac{L_{2,n}}{\alpha_{1,n}^3\alpha_{2,n}^3}+\cdots+\frac{L_{J-1,n}}{\alpha_{1,n}^3\alpha_{2,n}^3\cdots\alpha_{J-1,n}^3}~,
\end{equation}
and 
\begin{equation}\label{eq:contraction_factor_attractor}
    \alpha_{n}=\alpha_{1,n}\alpha_{2,n}\cdots\alpha_{J,n} ~.
\end{equation}
Similarly, we obtain 
\be 
C_{0,n+1} \equiv C_{J,n} = C_{0,n} \alpha _n^{1/4} e^{\mi \varphi},
\label{eq:amplitude_nth_loop}
\ee
with 
\begin{equation}\label{eq:phase_shift_attractor}
    \varphi=\varphi_1+\varphi_2+\cdots+\varphi_J.
\end{equation}
Note that $\alpha_{J,n}=\alpha_{0,n+1}$ and $\varphi_J=\varphi_0$.
For the first segment of the first cycle, the source is at $P_{0,1}$, so $L_{0,1}=0$ and the amplitude $C_{0,1}$ is given by the expression (\ref{eq:critical_latitude_C0}) of $C_0$. 

Although a given parameter $\alpha_{j,n}$ can be larger than 1, the product (\ref{eq:contraction_factor_attractor}) that defines $\alpha_n$ 
is necessarily smaller than 1 (for $n$ sufficiently large) because the critical ray converges toward an attractor. 
Its limit value $\alpha_{\infty}$ corresponds to the contraction factor of the attractor. 
The amplitude of the beam therefore goes rapidly to zero as  one gets close to the attractor. 
This guarantees that, although the various contributions superimpose on each other close to the attractor, the sum will remain finite on the attractor.   
The expression obtained by summing all the contributions coming from the segments $(0,n)$ with $n$ ranging from 1 to $\infty$ 
is then well defined. It can be written 
as    
\begin{equation}\label{eq:critical-latitude solution}
    v_\parallel\sim \sum_{n=1}^{\infty}v_{\parallel (0,n)} \quad , \quad \psi\sim\sum_{n=1}^{\infty}\psi_{0,n}
\end{equation}
for the parallel velocity  and the streamfunction respectively. 
 These expressions are expected to provide an asymptotic solution close to segments $(0,n)$. In the following, they will be referred to as the critical-latitude solution. 
In the next section, they are plotted and compared to numerical solutions. 

\subsection{Results} \label{sec:results}

The numerical solutions are obtained for Ekman numbers as low as $10^{-11}$ for which the scale separation between the wave beams and the domain size is clear. 
For simplicity, the velocity components $v_\phi$ in 3D and $v_y$ in 2D are used for comparison.
Other velocity components follow a similar trend. 

\begin{figure}
\centering
\begin{subfigure}{.5\textwidth}
 \centering
  \includegraphics[width=\textwidth]{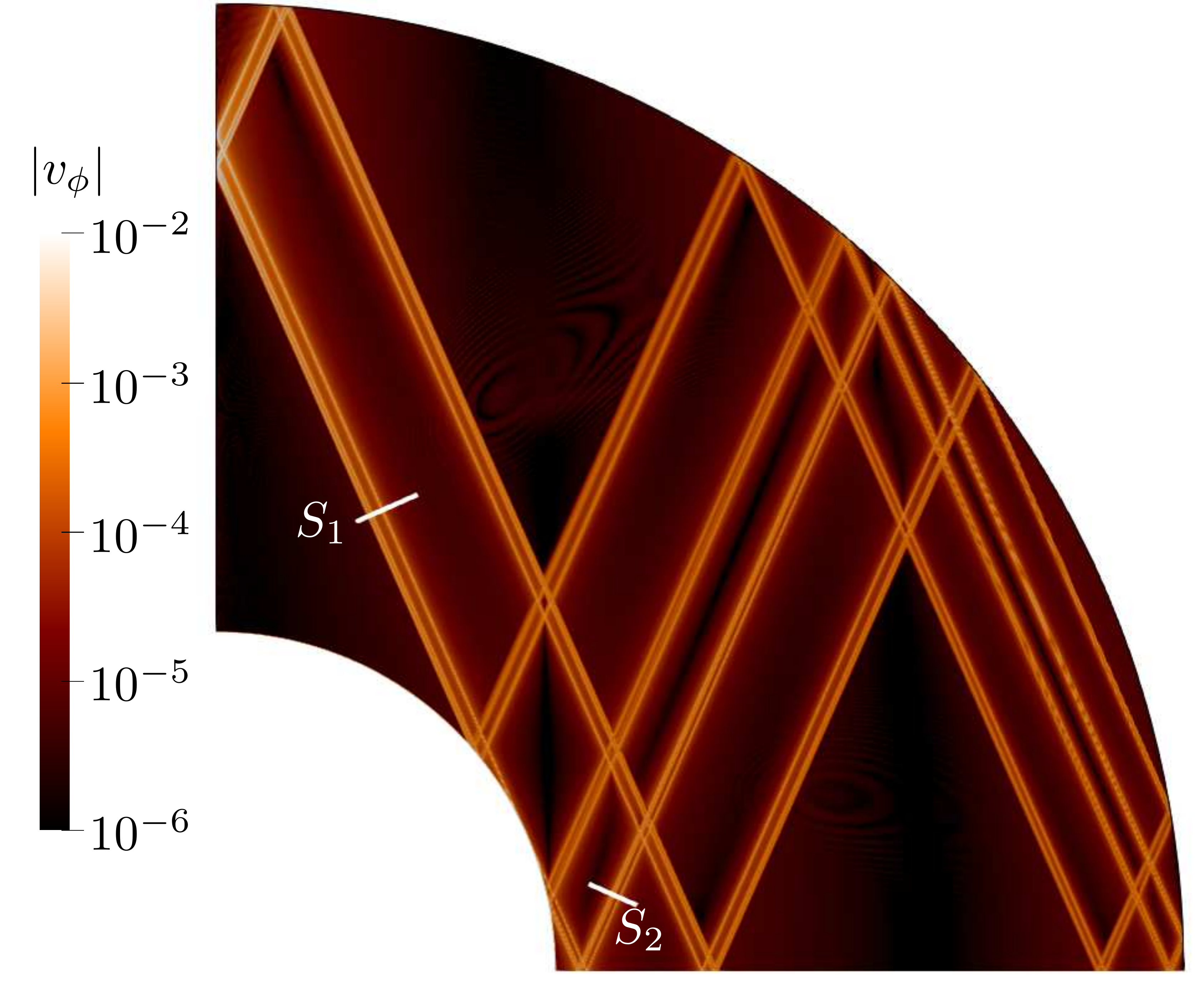}
  \caption{ Numerical result of the amplitude of $v_\phi$ at $E=10^{-11}$.}
  \label{fig:eta=0.350_omega=0.8102_contour}
\end{subfigure}%
\begin{subfigure}{.5\textwidth}
\centering
  \includegraphics[width=1\textwidth]{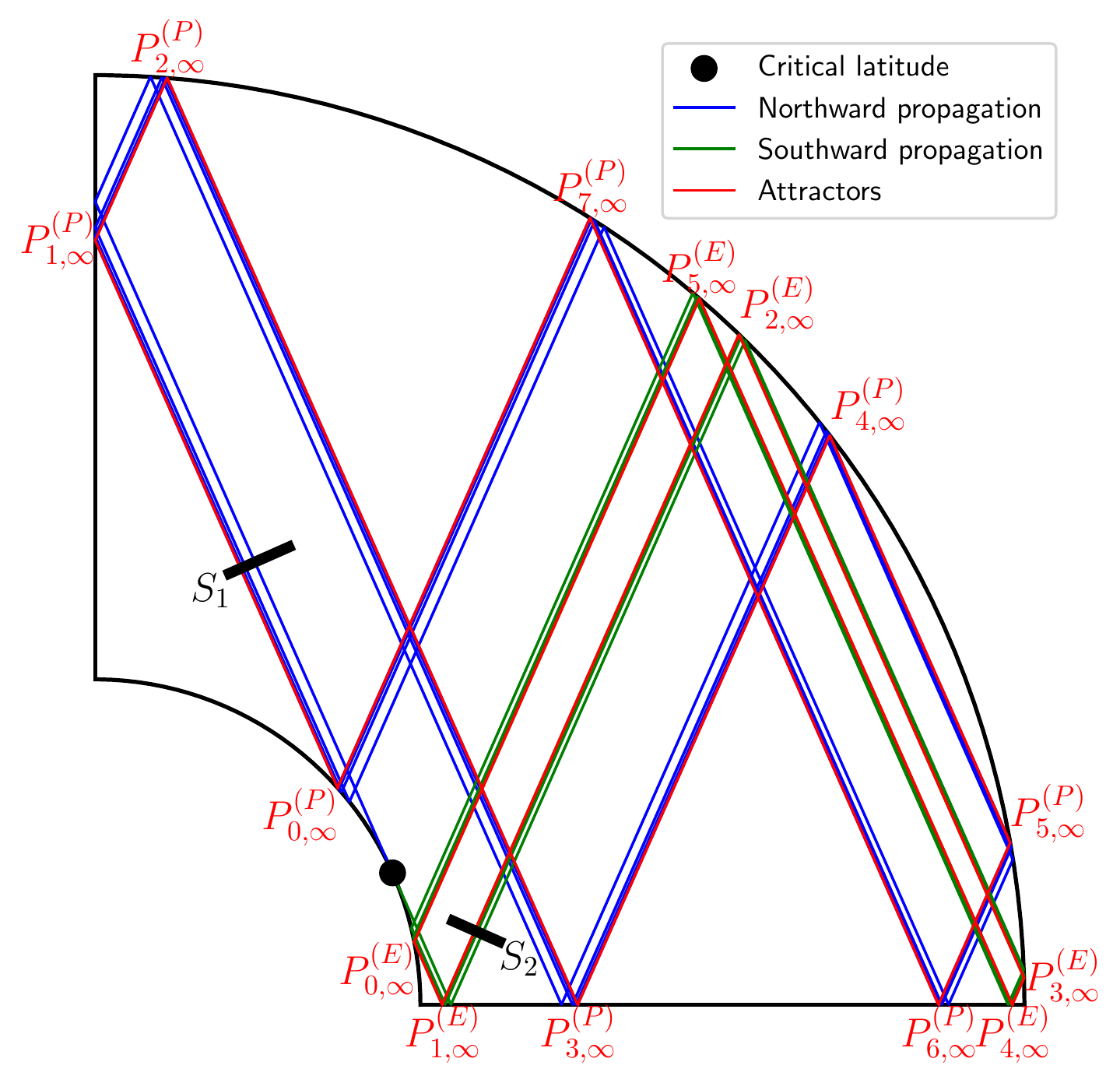}
  \caption{Ray paths from the critical latitude on the inner core.}
  \label{fig:eta=0.350_omega=0.8102_ray_paths}
\end{subfigure}
\caption{Results corresponding to the 3D libration of the inner core for $\eta=0.35$ and $\omega=0.8102$.}
\label{fig:eta=0.350_omega=0.8102}
\end{figure}

We consider the wave pattern with two coexisting attractors in a spherical shell as discussed by \citet{tilgnerDrivenInertialOscillations1999} and \citet{rieutordInertialWavesRotating2001}.
The aspect ratio and the frequency for this case are $\eta=0.35$ and $\omega=0.8102$ respectively.
The forcing is imposed on the inner core.
The numerical result of the 3D libration at $E=10^{-11}$ illustrated by the amplitude of $v_\phi$ is shown in figure \ref{fig:eta=0.350_omega=0.8102_contour}. %, where two coexisting attractors are observed. 
The wave pattern is consistent with the ray paths from the critical latitude on the forced inner core (see figure \ref{fig:eta=0.350_omega=0.8102_ray_paths}). 
The ray propagating northward from the critical latitude (in blue color) converges onto the polar attractor $P_{0,\infty}^{(P)}\cdots P_{7,\infty}^{(P)}$, while that propagating southward from the critical latitude (in green color) converges onto the equatorial attractor $P_{0,\infty}^{(E)}\cdots P_{5,\infty}^{(E)}$. 
The corresponding 2D results are not shown because the wave pattern and the ray paths are the same for the same aspect ratio and frequency. 
However, one should note that the phase shift $\varphi$ varies for different attractors and forcings. For the polar attractor with one vertex on the axis $Oz$, the phase shifts are $\pi/2$, $\pi$ and $0$ for the 3D libration, 2D symmetric and antisymmetric forcings, respectively. For the equatorial attractor, there is no phase shift for any of the forcing as this attractor does not touch the axis $Oz$.  

\begin{figure}
    {\centering
    \includegraphics[width=\textwidth]{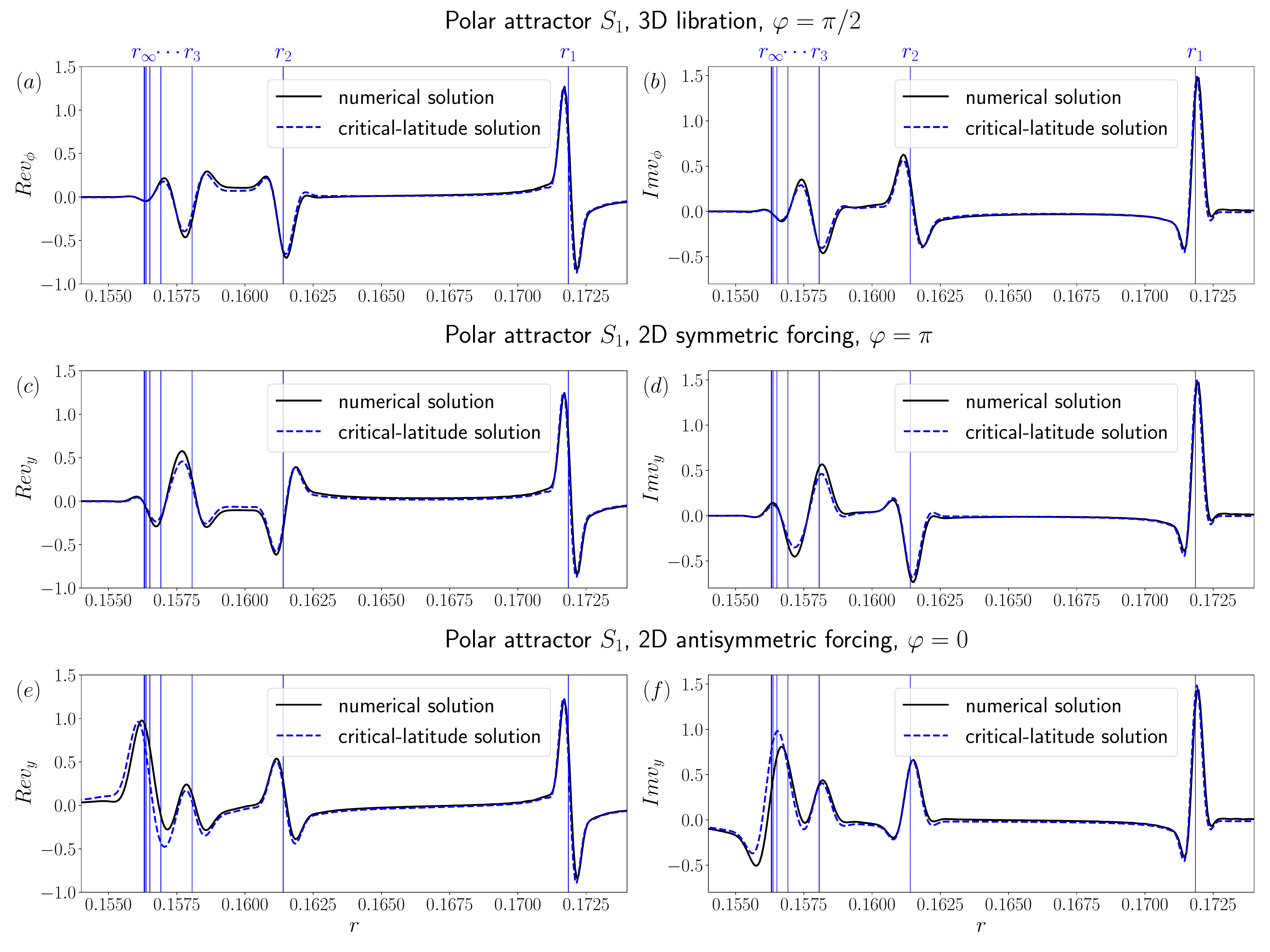}}
    \caption{Comparison of velocity profiles between the critical-latitude asymptotic solutions and the numerical solutions on the cut $S_1$ of the polar attractor shown in figure \ref{fig:eta=0.350_omega=0.8102} at $E=10^{-11}$ for three forcings: ($ab$) 3D libration (phase shift $\varphi=\pi/2$); ($cd$) 2D symmetric forcing (phase shift $\varphi=\pi$); ($ef$) 2D antisymmetric forcing (no phase shift). ($ace$) are the real parts; ($bdf$) are the imaginary parts.}
    \label{fig:eta=0.350_omega=0.8102_i=1_1e-11_cut}
\end{figure}

\begin{figure}
    \centering
    \includegraphics[width=\textwidth]{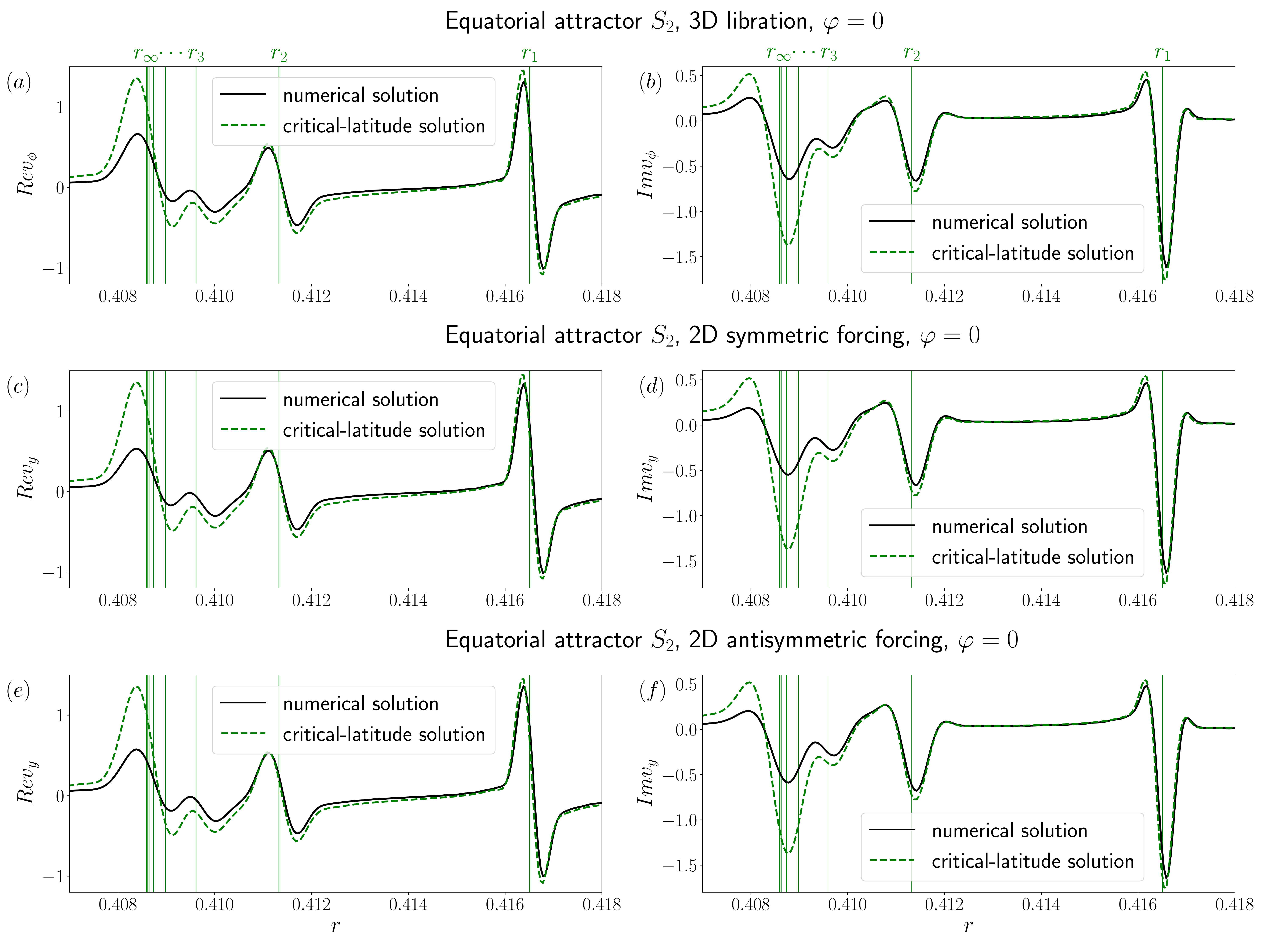}
    \caption{Same caption as in figure \ref{fig:eta=0.350_omega=0.8102_i=1_1e-11_cut} but on the cut $S_2$ of the equatorial attractor. There is no phase shift for all the three forcings.}
    \label{fig:eta=0.350_omega=0.8102_i=2_1e-11_cut}
\end{figure}

Two cuts crossing the two attractors are chosen in order to validate the critical-latitude asymptotic solution given by equation~\eqref{eq:critical-latitude solution}.
Figures \ref{fig:eta=0.350_omega=0.8102_i=1_1e-11_cut} and \ref{fig:eta=0.350_omega=0.8102_i=2_1e-11_cut} compare the velocity profiles between the asymptotic solutions and the numerical solutions at $E=10^{-11}$ on the cuts $S_1$ and $S_2$ respectively (see figure \ref{fig:eta=0.350_omega=0.8102}).  
The cut $S_1$ on the polar attractor is only crossed by the ray propagating northward from the critical latitude (blue lines in figure \ref{fig:eta=0.350_omega=0.8102_ray_paths}), while the cut $S_2$ on the equatorial attractor is only crossed by the ray propagating southward (green lines in figure \ref{fig:eta=0.350_omega=0.8102_ray_paths}).
In figures \ref{fig:eta=0.350_omega=0.8102_i=1_1e-11_cut} and \ref{fig:eta=0.350_omega=0.8102_i=2_1e-11_cut}, the vertical lines show the positions of the northward and southward critical rays when they cross $S_1$ and $S_2$ respectively.
These critical positions correspond to different successive loops.
From the rightmost critical position ($r_1$) to the leftmost one ($r_\infty$), the critical ray propagates from the first loop ($n=1$) to the final loop ($n=\infty$) and from the critical latitude to the final attractor.
The critical-latitude solutions on these two cuts $S_1$ and $S_2$ are built by propagating the self-similar solutions from the critical latitude northward and southward respectively, by using the infinite sum of self-similar solutions (\ref{eq:critical-latitude solution}); see the dashed lines in figures \ref{fig:eta=0.350_omega=0.8102_i=1_1e-11_cut} and \ref{fig:eta=0.350_omega=0.8102_i=2_1e-11_cut} respectively. 
Since the amplitude decreases exponentially the summation is conducted to a large enough number of loops in order to ensure convergence (around 150 loops in practice). 
The amplitudes are rescaled according to table \ref{tab:absolute_value_of_complex_amplitude}, in order to make sure that the wave beams from the critical latitude possess the same amplitudes for all the three forcings. 
Note that the radial dependence of the 3D configuration is removed by multiplying the velocity with $\sqrt{r}$.

Around the first critical position $r_1$, the wave beam from the critical latitude are within the first loop and have not experienced any contraction or expansion on the boundaries. 
It takes the same shape for all the three forcings and the self-similar solution agrees with the numerical solution very well for both two cuts and all the three forcings, as shown by the profiles around $r_1$ in figures \ref{fig:eta=0.350_omega=0.8102_i=1_1e-11_cut} and \ref{fig:eta=0.350_omega=0.8102_i=2_1e-11_cut}. 
After one loop, the wave beam moves on to the next critical position $r_2$. The amplitude decreases as expected. The agreement between the self-similar solution and the numerical solution is still good; see the profiles around $r_2$ in figures \ref{fig:eta=0.350_omega=0.8102_i=1_1e-11_cut} and \ref{fig:eta=0.350_omega=0.8102_i=2_1e-11_cut}.
However, its shape is now dependent on the phase shift it has experienced during the first loop.
For the polar attractor on $S_1$, the profile around $r_2$  changes compared to that around $r_1$ for the 3D libration and 2D symmetric forcing since there is a nonzero phase shift for both these cases (see figures \ref{fig:eta=0.350_omega=0.8102_i=1_1e-11_cut}$(abcd)$). 
The profiles between $r_1$ and $r_2$ remain similar for the 2D antisymmetric forcing since there is no phase shift (figure \ref{fig:eta=0.350_omega=0.8102_i=1_1e-11_cut}$(ef)$).
For the equatorial attractor on $S_2$, there is no phase shift and the profiles remain similar from $r_1$ to $r_2$ as shown in figure \ref{fig:eta=0.350_omega=0.8102_i=2_1e-11_cut} for all the three forcings.
A similar behaviour can be observed from the critical position $r_2$ to the next position $r_3$.
Interestingly, the equatorial attractor profiles on $S_2$, which do not cross the rotation axis and therefore do not alter the phase, are almost the same for all forcing types.

When the wave beam moves on to the position of the attractor ($r_\infty$), successive critical positions become very close to each other and the profiles from different loops are not well separated. 
Finally, the wave beam just propagates on the attractor and the summation of the self-similar solutions is conducted there.
As shown by the profiles around the positions of the attractors $r_\infty$ in figures \ref{fig:eta=0.350_omega=0.8102_i=1_1e-11_cut} and \ref{fig:eta=0.350_omega=0.8102_i=2_1e-11_cut}, the attractor with a nonzero phase shift is weak (figure \ref{fig:eta=0.350_omega=0.8102_i=1_1e-11_cut}$(abcd)$), while that without phase shift is strong (figures \ref{fig:eta=0.350_omega=0.8102_i=1_1e-11_cut}$(ef)$ and \ref{fig:eta=0.350_omega=0.8102_i=2_1e-11_cut}). 
This phenomenon can be explained by the summation of the self-similar solutions on the attractor.
When there is a phase shift on the attractor path, the self-similar solutions of successive loops with different phases cancel out, which makes the local solution in the vicinity of the attractor negligible after summation.
Otherwise, the self-similar solutions with the same phase accumulate on the attractor, which makes the solution much stronger.
For the weak polar attractor ($r_\infty$ in figure \ref{fig:eta=0.350_omega=0.8102_i=1_1e-11_cut}$(abcd)$), the critical-latitude solution \eqref{eq:critical-latitude solution} is consistent with the numerical solution.
However, for the strong polar attractor ($r_\infty$ in figure \ref{fig:eta=0.350_omega=0.8102_i=1_1e-11_cut}$(ef)$), the asymptotic solution does not perform as well as for the positions far from the attractor.
This deviation is much more obvious for the strong equatorial attractor ($r_\infty$ in figure \ref{fig:eta=0.350_omega=0.8102_i=2_1e-11_cut}), where the amplitudes are largely overestimated and the critical-latitude solution (\ref{eq:critical-latitude solution}) deviates from the numerical solutions gradually as the ray converges towards the attractor.

\begin{figure}
    \centering
    \includegraphics[width=1.0\textwidth]{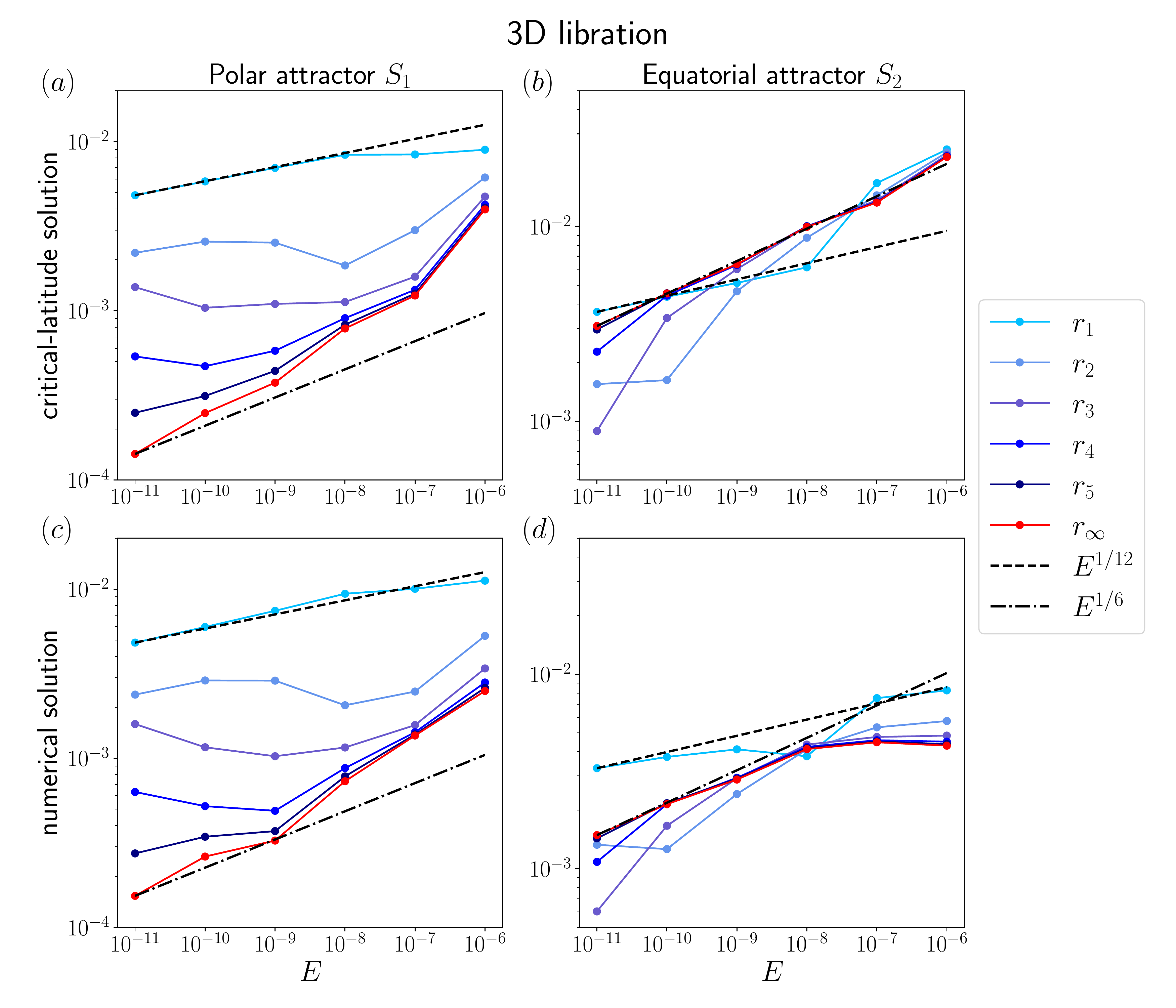}
    \caption{Velocity amplitude scalings  of the critical-latitude solution ($ab$) and the numerical solution ($cd$) at the critical positions on the cuts $S_1$ ($ac$) and $S_2$ ($bd$) for the 3D libration and at the Ekman numbers $[10^{-11}, 10^{-6}]$. $r_1$, $r_2$, $\cdots$ and $r_\infty$ are the critical positions shown in figures \ref{fig:eta=0.350_omega=0.8102_i=1_1e-11_cut} and \ref{fig:eta=0.350_omega=0.8102_i=2_1e-11_cut}.}
    \label{fig:amplitude_scalings}
\end{figure}

In order to investigate what is happening around the attractor, the velocity amplitude scalings with Ekman number of both the critical-latitude asymptotic solution \eqref{eq:critical-latitude solution} and the numerical solution at the critical positions on the cuts $S_1$ and $S_2$ are shown in figure \ref{fig:amplitude_scalings} for the 3D libration. 
Similar behaviour can be observed for the other two forcings and are not shown here.
The critical-latitude solution (\ref{eq:critical-latitude solution}) and the numerical solution at the first critical position $r_1$ follow the expected scaling $E^{1/12}$, which validates the Ekman number scaling of the wave beams from the critical latitude for a libration frequency different from that in HFRL22. 
Then the amplitude on $S_1$ (weak polar attractor with phase shift) decreases to a weaker level as the ray gets closer to the position of the attractor $r_\infty$ and the scaling is eventually closer to $E^{1/6}$. 
The scaling $E^{1/6}$ around $r_\infty$ is more obvious for the other cut $S_2$ of the strong equatorial attractor without phase shift. 
However, the corresponding prefactor is over-predicted by the critical-latitude solution.

The change of scaling of the critical-latitude solution could have been anticipated from expression (\ref{eq:amplitude_nth_loop}) for its amplitude. We have seen 
that because $\alpha_n <1$, the amplitude of the self-similar beam decreases as it gets closer to the attractor. But  the beam has also a finite width of order $E^{1/3}$, 
so the contributions obtained from each cycle superimpose on each other when the critical ray gets at a distance of this order from the attractor.  This stops the decreasing of the 
amplitude after a number $n_s$ of cycles that can be approximately estimated by $(\log E^{1/3})/(\log \alpha_\infty)$ which corresponds to the number of contraction needed to go from 1 
to $E^{1/3} $ with the contraction factor $\alpha_\infty$.  
The amplitude $C_{0,n_s}$ has then decreased from its initial value $C_0$ by a factor $\alpha_1^{1/4} \alpha_2^{1/4} \cdots \alpha_{n_s}^{1/4}$, which is close to $\alpha_{\infty}^{n_s/4}\approx E^{1/12}$ .  
The velocity amplitude of the critical-latitude solution is therefore expected to become $O(E^{1/12})$ smaller  close to the attractor and therefore of order $E^{1/6}$, as observed.

The amplitude of the streamfunction of the critical-latitude solution also decreases from $O(E^{5/12})$ to $O(E^{1/2})$ as we get close to the attractor. 
Then, it becomes of the order of the Ekman pumping   (see Appendix~\ref{app:Ekman_pumping}). 
This means that the hypothesis of no Ekman pumping that has been used to obtain the reflection laws of the beam in \S \ref{sec:reflections} breaks down. 
In particular,  relation (\ref{exp:reflection-psi}) should not be valid close to the attractor.  
We suspect that the discrepancies observed close to the attractor between the critical-latitude solution and the numerical solution  
are due to this effect. 

In the next section, we develop a new asymptotic theory to describe the solution close to the attractor. This theory, that takes into account the 
Ekman pumping close to the attractor, is based on ideas originally developed by O05.

\section{Solution close to the attractors}\label{sec:solution_of_attractor}

\subsection{Asymptotic theory}
\label{sec:asym_attractor}

In order to apply the results of O05, we first consider 2D configurations without phase shift or reflection on the axis.
In this framework, the inviscid problem can be solved using the 2D streamfunction only and a global solution for the streamfunction is obtained 
 as a sum of two functions which are constant along the characteristics of the problem \cite[e.g.][]{maasGeometricFocusingInternal1995}.

In the present work, we are looking for a solution  localised near an attractor, say $P_{0,\infty}\cdots P_{J-1,\infty}$ as illustrated in figure \ref{fig:schmatic_of_propagation},
which can be written as a sum of local solutions  $ \psi_{j,\infty}^{(2D)}(x_{\perp j,\infty})$ valid close to the segment $(j,\infty)$ only.  
In the 2D inviscid framework, these local solutions just mean that the solution is transported from $P_{j,\infty}$ to $P_{j+1 \infty}$  along the lines $x_{\perp j,\infty} = Cst$ without modification. 
Whereas the critical-latitude solution was reflected on boundary without modification, the solution constructed by O05 is directly forced by the boundary condition in the neighborhood of the attractor.
Close to a point $P_{j\infty}$, the solution is expected to be composed of the incident solution  $ \psi_{j,\infty}^{(2D)}(x_{\perp j,\infty})$ and of the reflected solution $ \psi_{j+1,\infty}^{(2D)}(x_{\perp j+1,\infty})$
and satisfies the boundary condition on the surface :
\be
\psi_{j,\infty}^{(2D)}(x_{\perp j,\infty}) +  \psi_{j+1,\infty}^{(2D)}(x_{\perp j+1,\infty}) = \psi_{j, \infty}^{(EP)}  ~,~   
\label{eq:reflection_attractor}
\ee
where the condition of being on the surface  close to $P_{j,\infty}$ means that
\be 
x_{\perp j+1,\infty} = \alpha _{j,\infty} x_{\perp, j,\infty}.
\ee
In (\ref{eq:reflection_attractor}), $\psi_{j,\infty}^{(EP)}$ is the value of the streamfunction prescribed at $P_{j,\infty}$.
In our case, this prescribed value is given by the Ekman pumping at the surface (or is zero if there is no Ekman pumping).  

If we apply these conditions at each point $P_{0,\infty}, P_{1,\infty}  \cdots P_{J-1,\infty}$ of the attractor, we end up, after a complete cycle, with 
an equation for each $\psi_{j,\infty}$ which can be written as 
\be
\psi_{j,\infty}^{(2D)}(\alpha_\infty  x_{\perp j,\infty}) - \psi_{j,\infty}^{(2D)}(x_{\perp j,\infty})  =\epsilon_{j,\infty} \delta,
\label{eq:attractor2D}
\ee
with 
\be
\delta = \sum_{k=1}^{J/2}( \psi_{2k, \infty}^{(EP)} -\psi_{2k+1, \infty}^{(EP)}   )  ~,
\label{eq:delta_definition_2D}
\ee
and 
\be
\alpha_\infty = \alpha_{1,\infty} \alpha_{2,\infty} \cdots \alpha_{J,\infty},
\ee
 where we have used the fact that the point $P_{j+J,\infty}$ corresponds to the point $P_{j,\infty}$ (implying that  $\psi_{j,\infty}=\psi_{j+J,\infty}$ and $\psi_{j+J, \infty}^{(EP)} = \psi_{j, \infty}^{(EP)} $ ). 
The parameter $\epsilon_{j,\infty}$ is the sign in the streamfunction definition (\ref{eq:local_velocity_streamfunction_transform}).
 The parameter $\alpha_\infty$ is the contraction factor of the attractor. 
Equation (\ref{eq:attractor2D}) is a functional constraint on the inviscid solution close to a 2D attractor.  It is identical to the equation (3.17) of O05.  

The general solution of this equation can be written as 
\begin{equation}\label{eq:O05_solution}
   \psi_{j,\infty}^{(2D)}(x_{\perp j,\infty})  =  \frac{\epsilon_{j,\infty} \delta}{\ln\alpha_\infty}\ln |x_{\perp j,\infty}| +\sum_{n=-\infty}^{+\infty}h_n^{\pm} |x_{\perp j,\infty}|^{\frac{2n\pi\mathrm{i}}{\ln \alpha_\infty}} ~, 
\end{equation}
where the $\pm$ sign is for positive or negative $x_{\perp j,\infty}$. 
Interestingly, the dominant logarithmic part of this solution has a simple expression that only depends on the contraction factor $\alpha_\infty$ and the forcing term $\delta$. 
This part corresponds to a particular solution of (\ref{eq:attractor2D}) while the sum is a general homogeneous solution. 
Contrarily to O05, we shall not try to determine this homogeneous solution.  We shall only keep the dominant logarithmic term to 
describe the inviscid solution close to the attractor. This hypothesis is not  justified from an  asymptotical point of view but O05  showed 
that the correction associated with the homogeneous part was very small for his case. 

 If  we  only keep the particular solution, we then get  a simple inviscid expression for the  parallel velocity which is
\be
v_{\parallel j,\infty}^{(2D)} \sim   \frac{\delta}{\ln\alpha_\infty} x_{\perp j,\infty}^{-1} .
\label{exp:vparallel}
\ee
As already explained above, this singular behavior can be smoothed by viscosity by introducing  the self-similar solution of Moore \& Saffman. The 
viscous solution that matches with the singular  behavior (\ref{exp:vparallel}) is 
\be
v_{\parallel j,\infty}^{(2D)} \sim   C_0^{(A)} H_1(x_{\parallel j,\infty}, x_{\perp j,\infty})  ,
\label{exp:vattractor2D}
\ee
with 
\be 
C_0^{(A)} = \frac{\delta}{\ln\alpha_\infty} E^{-1/3} ,
\label{exp:C0A}
\ee
where $H_m(x_\parallel,x_\perp)$ has been defined in (\ref{eq:self_similar_solution}). 

In the above expression, the virtual source of the beam, that is the position where  $x_{\parallel j,\infty} =0 $ is however not known. 
This position can be obtained by using the argument developed in \S \ref{sec:propagation} for the critical-latitude solution.
In particular, if as above, $x_{\parallel j,\infty}$ is written as
\be 
x_{\parallel j,\infty} = L_{j,\infty}^{(s)} +  x_{\parallel j,\infty}' 
\label{exp:xparallel}
\ee 
with $x_{\parallel j,\infty}'=0$ at $P_{j,\infty}$, the distance  $L_{j,\infty}^{(s)}$ satisfies the equations (\ref{exp:Ljn}) and (\ref{exp:L0(s)}) with $n\rightarrow\infty$.
For the first segment between $P_{0, \infty}$ and $P_{1, \infty}$, we then get using  (\ref{exp:L0(s)}) 
\be
L_{0,\infty} ^{(s)} = (L _{0,\infty}^{(s)} + \Lambda_\infty )\alpha_\infty^3 ,
\ee
that is 
\begin{equation}\label{eq:attractor_source}
  L_{0,\infty} ^{(s)}  =\Lambda_\infty \frac{\alpha_\infty^3}{1-\alpha_\infty^3},
\end{equation}
where $\Lambda_\infty$ and $\alpha_\infty$ are given by (\ref{eq:length}) and (\ref{eq:contraction_factor_attractor}) with $n \rightarrow \infty$. 

It is worth noting that the amplitude of the attractor solution does not change along the cycle. This is due to the particular value of the index of $m$ of the self-similar solution, that is $m=1$ for the attractor solution, 
which guarantees that the amplitude does not change when the beam reflects on the boundary, as prescribed by the reflection law 
(\ref{eq:reflection_laws}b). 
  
For the 3D configurations or when there is a phase shift during a cycle, the above considerations have to be modified.
First note that in 3D axisymmetric geometries, the streamfunction is not identically propagated along characteristics  as it is in 2D. It evolves according to a propagator defined by the Riemann function, which is a Legendre function of index -1/2 for axisymmetric solutions considered here \cite[but see \S2.3.2 of][for more details]{rieutordInertialWavesRotating2001}.  
In other words, there is no simple global expression of the inviscid solution for the streamfunction in 3D. 
However, if the solution varies on small scales compared to the distance to the axis, the propagation is almost as in 2D: in that case, an approximate local solution can be 
obtained far from the axis in the form 
\be
   \psi=\sqrt{r}\tilde{\psi}(x_\perp) ~ ,~~ v_\parallel  = \frac{\tilde{v}_\parallel (x_\perp)}{\sqrt{r}},
   \label{exp:local3D}
\ee
where the $\sqrt{r}$ factor guarantees that these approximations are valid up to second order corrections. 
The same analysis as above can then be performed for $\tilde{\psi}$ as long as we are far from the axis. 
This leads to a 3D expression for the local solution near an attractor without phase shift 
which is directly obtained from (\ref{exp:vattractor2D}) as
\be
v_{\parallel j,\infty} \sim  \frac{C_0^{(A)}}{\sqrt{r}} H_1(x_{\parallel j,\infty}, x_{\perp j,\infty})  , 
\label{exp:vattractor3D}
\ee
with $C_0^{(A)}$ given by (\ref{exp:C0A}) but with an expression of  $\delta$ slightly different 
\be
\delta = \sum_{k=1}^{J/2}\left( \frac{\psi_{2k, \infty}^{(EP)}}{\sqrt{r_{2k,\infty}}} -\frac{\psi_{2k+1, \infty}^{(EP)}}{\sqrt{r_{2k+1,\infty}}}\right)  .
\label{eq:delta_definition_3D}
\ee
For the solution along the first segment $P_{0, \infty}P_{1, \infty}$, $x_{\parallel 0,\infty}$ is still defined by (\ref{exp:xparallel}) and (\ref{eq:attractor_source}).

For the attractor without reflection on the axis, it is the expressions (\ref{exp:vattractor2D}) for 2D configurations and   (\ref{exp:vattractor3D}) for 3D configurations that we shall use and compare to our numerical data.

We now want to consider the case of an attractor touching the axis.
For the critical-latitude solution, we have seen that a phase shift could be generated as the beam reflects on the axis. 
A similar phenomenon is expected for the attractor solution. 
Let us first consider the 2D configurations. We have seen that in that case, the presence of a phase shift depends on the symmetry of the forcing with respect to the $z$ axis.
A phase shift of  $\varphi=\pi$ is expected for a symmetric forcing, while no phase shift is present for an antisymmetric forcing. 
This is easily understood as changing $x$ into $-x$ changes $x_{\perp 0,\infty}$ into $x_{\perp 1,\infty}$. 
The local solutions $\psi_{0,\infty}^{(2D)}(x_{\perp 0,\infty})$ and $\psi_{1,\infty}^{(2D)} (x_{\perp 1,\infty})$, valid close to the two lines $x_{\perp 0,\infty}=0$ and $x_{\perp 1,\infty}=0$ respectively,
should therefore satisfy the same symmetry as the forcing, that is,
\be
\psi_{1,\infty}^{(2D)} (x_{\perp 1,\infty}) = \psi_{0,\infty}^{(2D)}(x_{\perp 0,\infty})  
\label{eq:reflectionsym}
\ee
for the symmetric forcing
and 
\be
\psi_{1,\infty}^{(2D)} (x_{\perp 1,\infty}) = - \psi_{0,\infty}^{(2D)}(x_{\perp 0,\infty})  
\label{eq:reflectionanti}
\ee
for the antisymmetric forcing, when $x_{\perp 0,\infty}=x_{\perp 1,\infty}$.   
For the antisymmetric case, the axis is therefore as a vertical boundary without Ekman pumping [compare (\ref{eq:reflectionanti}) to (\ref{eq:reflection_attractor})]. The same solution as for a 2D attractor not refecting on the axis can therefore be used.
For the symmetric case, this is no longer the case as, owing to (\ref{eq:reflectionsym}), (\ref{eq:attractor2D}) now becomes
\be
\psi_{j,\infty}^{(2D)}(\alpha_\infty  x_{\perp j,\infty}) + \psi_{j,\infty}^{(2D)}(x_{\perp j,\infty})  =\epsilon_{j,\infty} \delta,
\label{eq:attractor2Danti}
\ee
if there is an even number of reflections on the axis. 
A particular solution to this equation is just  $ \psi_{j,\infty}^{(2D)} ( x_{\perp j,\infty}) = \epsilon_{j,\infty}\delta/2 $ so there is no  logarithmic singularity in the solution anymore. 
Thus, the inviscid expression (\ref{exp:vparallel}) is not obtained, neither its viscous counterpart (\ref{exp:vattractor2D}).

In 3D, as a phase shift of $\pi/2$ appears when the ray reflects on the axis, a similar phenomenon is expected if the number of 
reflections on the axis is not a multiple of 4. In that case, no logarithmic singularity should be present in the function $\tilde{\psi}_{j,\infty}$, and the analysis 
performed above should also break down.  
A weaker attractor solution is probably obtained in that case which could explain why no significant attractor contribution was observed in the numerical solution when there is a phase shift. 
%This could explain the observations made in the previous section: no contribution from an attractor solution was observed when there is an phase shift along the attractor. 
Finding the correct asymptotic form of the attractor solution in the presence of a phase shift is not an easy task. We 
leave it for future studies, probably in a simpler geometry.  

\subsection{Results}
\label{sec:results_attractor}

\begin{figure}
    \centering
    \includegraphics[width=\textwidth]{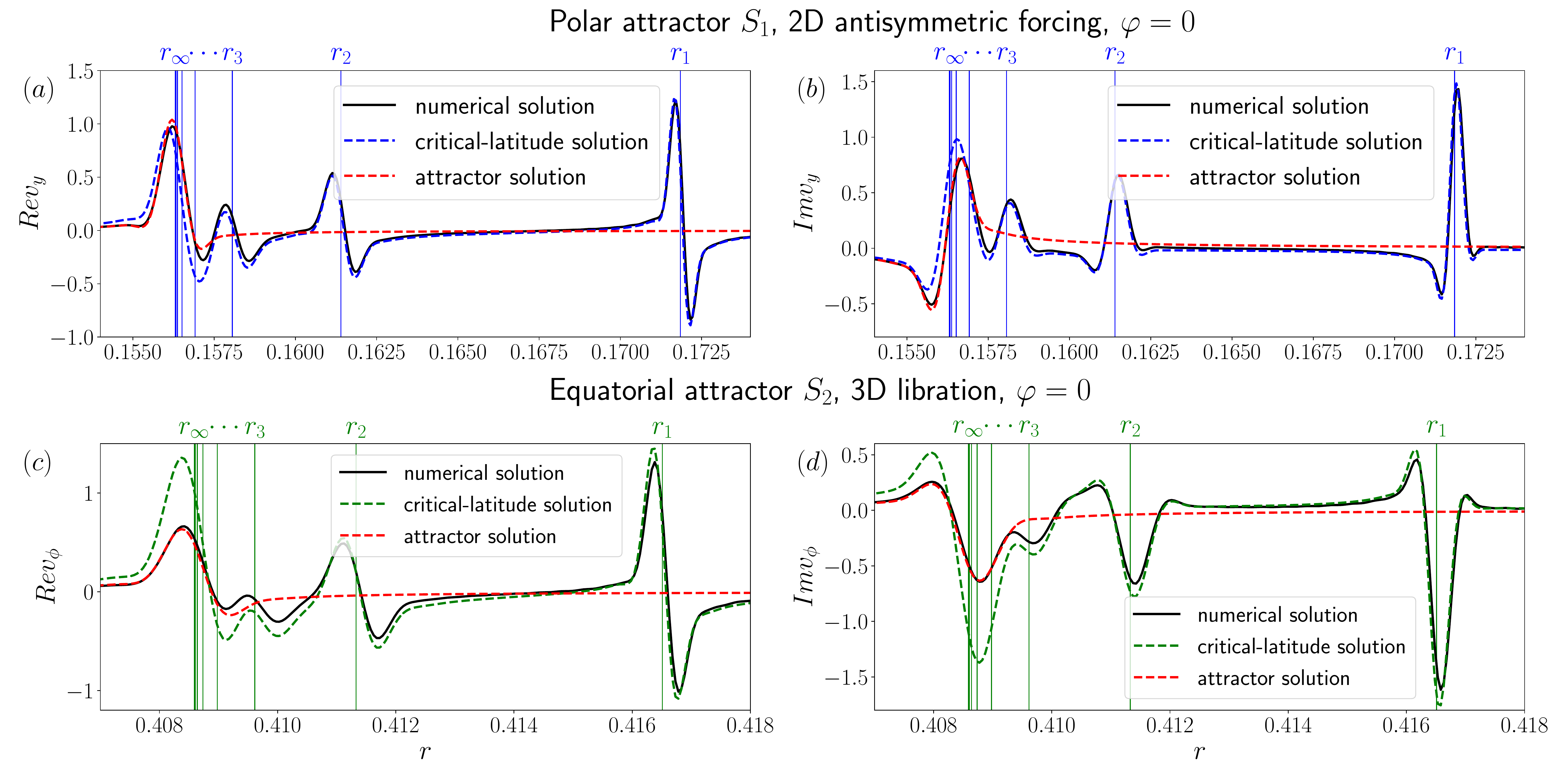}
    \caption{Comparison of velocity profiles between the numerical solution, the critical-latitude solution and the attractor solution at $E=10^{-11}$: $(ab)$ the cut $S_1$ of the polar attractor forced by the 2D antisymmetric forcing; $(cd)$ the cut $S_2$ of the equatorial attractor forced the 3D libration (the other 2D forcings for the equatorial attractor show the same results as $(cd)$).}
    \label{fig:attractor_profile}
\end{figure}

We now try to assess the performance of the attractor solution discussed above and see whether the rather poor performance of the critical-latitude solution \eqref{eq:critical-latitude solution} for the cases without phase shift (the polar attractor forced by the 2D antisymmetric forcing in figure \ref{fig:eta=0.350_omega=0.8102_i=1_1e-11_cut}$(ef)$ and the equatorial attractors in figure \ref{fig:eta=0.350_omega=0.8102_i=2_1e-11_cut}) is improved.
Since all the three forcings are imposed on the inner core, only one vertex for each attractor is forced, namely $P^{(P)}_{0,\infty}$ for the polar attractor and $P^{(E)}_{0,\infty}$ for the equatorial attractor (see figure \ref{fig:eta=0.350_omega=0.8102_ray_paths}). 
The forcing term $\delta$ (\ref{eq:delta_definition_2D}, \ref{eq:delta_definition_3D}) can be simplified to
\begin{equation}
  \delta = \left\{
    \begin{array}{ll}
      \psi^{(EP)}_{0,\infty}, & \mbox{2D}; \\[2pt]
       \psi^{(EP)}_{0,\infty}/\sqrt{r_{0,\infty}}, & \mbox{3D}.
  \end{array} \right.
\end{equation}
The values of the Ekman pumping at the positions $P^{(P)}_{0,\infty}$ and $P^{(E)}_{0,\infty}$ correspond to the formulae of the inner core in table \ref{tab:Ekman_pumping} of Appendix~\ref{app:Ekman_pumping}.
As shown in figure \ref{fig:attractor_profile}, the attractor solution (in red color) performs much better than the previous critical-latitude solution. 
It demonstrates the necessity of including the Ekman pumping into the asymptotic solution as one gets close to the attractor, especially for the strong attractors without phase shift.
However, it is unnecessary for the weak attractors with phase shift as shown by \ref{fig:eta=0.350_omega=0.8102_i=1_1e-11_cut}$(abcd)$, since there is no logarithmic singularity.

Note that we now face the problem of defining a transition between the critical-latitude solution valid during the first cycles and the attractor solution eventually valid close to the attractor.
While we do not discuss this aspect of the problem in this paper, further studies would be required to clarify this transition.

\begin{figure}
\centering
\begin{subfigure}{.5\textwidth}
 \centering
  \includegraphics[width=\textwidth]{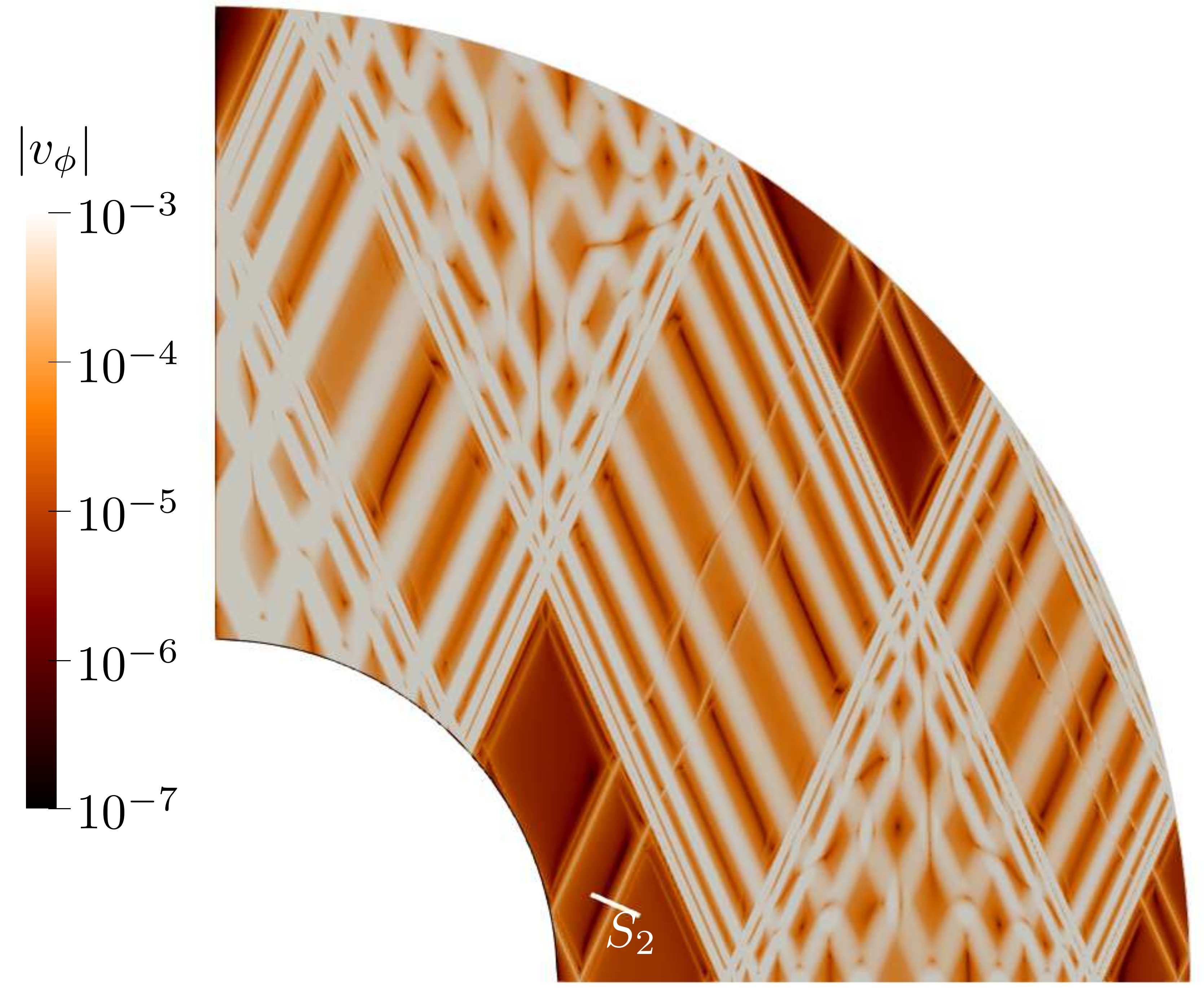}
  \caption{Numerical result of the amplitude of $v_\phi$ at $E=10^{-11}$.}
  \label{fig:eta=0.350_omega=0.8102_outercore_contour}
\end{subfigure}%
\begin{subfigure}{.5\textwidth}
\centering
  \includegraphics[width=1\textwidth]{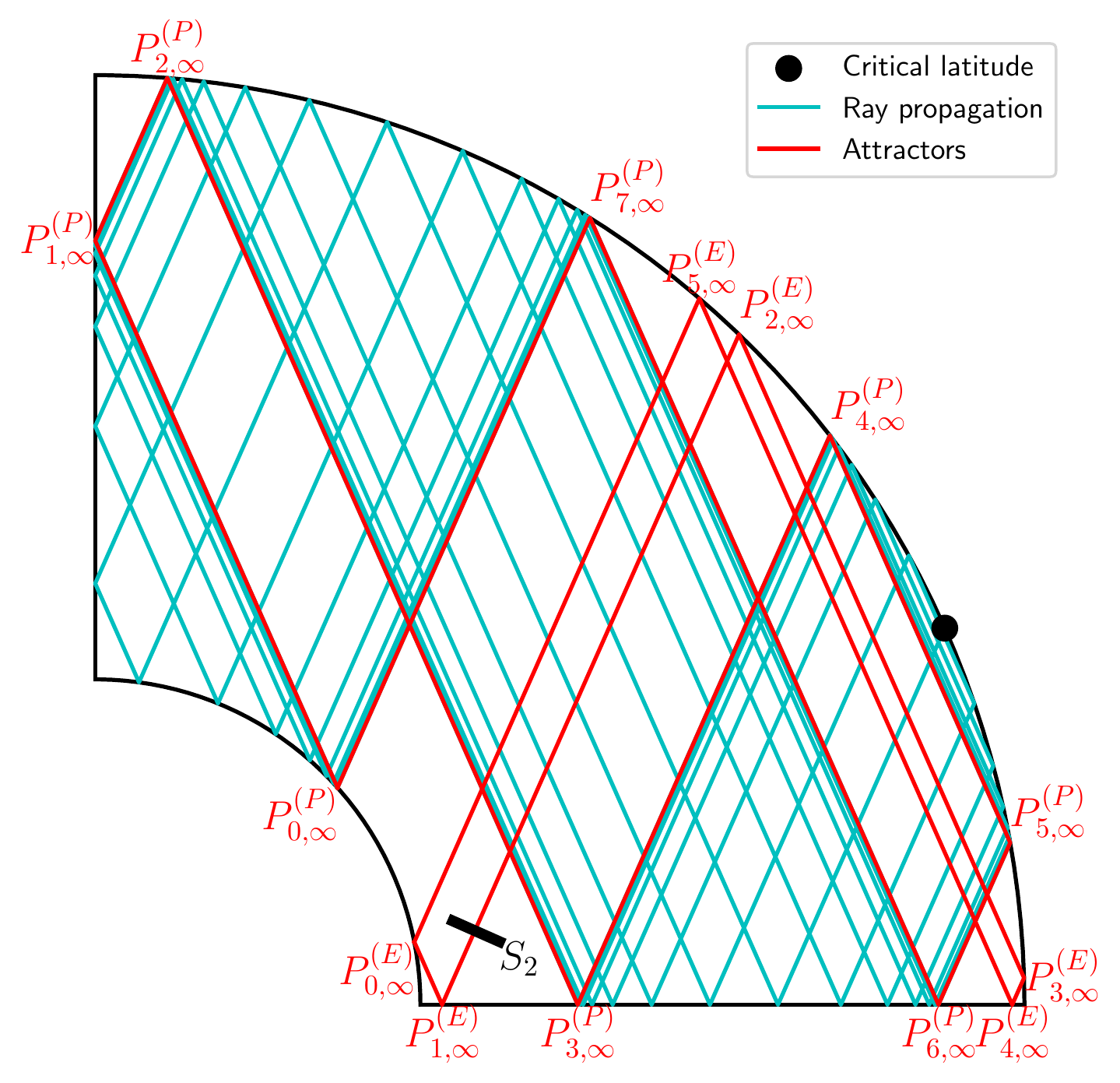}
  \caption{Ray paths from the critical latitude on the outer core.}
  \label{fig:eta=0.350_omega=0.8102_outercore_ray_paths}
\end{subfigure}
\caption{Results corresponding to the 3D libration of the outer core for $\eta=0.35$ and $\omega=0.8102$.}
\label{fig:eta=0.350_omega=0.8102_outercore}
\end{figure}

The forcings considered up to now were imposed on the inner core, which excites the wave beams from the critical latitude on the inner core. 
These wave beams propagate towards the attractors and coexist with them.  
In order to further validate the attractor solution, it is helpful to consider a configuration where the attractor is not affected by the propagation of the wave beams from the forced critical latitude. 
Fortunately, such configuration exists for the same aspect ratio and frequency but with the forcing imposed on the outer core.
The wave pattern of the 3D libration on the outer core is shown in figure \ref{fig:eta=0.350_omega=0.8102_outercore_contour}, which can be compared to the ray paths in figure \ref{fig:eta=0.350_omega=0.8102_outercore_ray_paths}. 
Since only the critical latitude on the outer core is forced, the only option for the initial propagation direction is pointing into the bulk. 
As shown in figure \ref{fig:eta=0.350_omega=0.8102_outercore_ray_paths}, this ray (in cyan color) propagates onto the polar attractor.
The corresponding wave beam from the critical latitude on the outer core should possess $E^{1/5}$ width and $E^{1/5}$ amplitude  \citep{kerswellInternalShearLayers1995,linLibrationdrivenInertialWaves2020},  but it will not be our concern here. 
More importantly, figure \ref{fig:eta=0.350_omega=0.8102_outercore_contour} shows that the equatorial attractor is still present although it is not connected to the ray emerging from the critical latitude on the outer core.
It should thus be directly forced by the Ekman pumping at the positions of the attractor on the outer core. 
The attractor solution can be built for this attractor since there is no phase shift associated with it.
Because the vertices $P_{2,\infty}^{(E)}$, $P_{3,\infty}^{(E)}$ and $P_{5,\infty}^{(E)}$ of the equatorial attractor are forced,
the forcing term $\delta$ (\ref{eq:delta_definition_2D}, \ref{eq:delta_definition_3D}) can be simplified to
\begin{equation}
  \delta = \left\{
    \begin{array}{ll}
      \psi^{(EP)}_{2, \infty}-\psi^{(EP)}_{3, \infty}-\psi^{(EP)}_{5, \infty}, & \mbox{2D}; \\[2pt]
      \psi^{(EP)}_{2, \infty}/\sqrt{r_{2,\infty}}-\psi^{(EP)}_{3, \infty}/\sqrt{r_{3,\infty}}-\psi^{(EP)}_{5, \infty}/\sqrt{r_{5,\infty}}, & \mbox{3D}.
  \end{array} \right.
\end{equation}
The values of the Ekman pumping at the positions $P_{2, \infty}^{(E)}$, $P_{3, \infty}^{(E)}$ and $P_{5, \infty}^{(E)}$ correspond to the formulae of the outer core in table \ref{tab:Ekman_pumping} of Appendix~\ref{app:Ekman_pumping}.
Figure \ref{fig:eta=0.350_omega=0.8102_outercore_profile} shows the comparison between the attractor solutions and the numerical solutions for the three forcings at three Ekman numbers. 
The amplitudes of the three forcings are rescaled. 
Good performance of the attractor solution is observed.
As the Ekman number decreases, the agreement between the two solutions becomes better.
The small ripples on the negative side of the similarity variable at the low Ekman numbers are wave beams from the critical latitude on the unforced inner core. 
They are much weaker and the accumulation of them on the attractor remain negligible compared to the attractor beam. 
Figure \ref{fig:eta=0.350_omega=0.8102_outercore_scaling} shows the Ekman number scalings of the attractor beams with a beam width and  a velocity amplitude in  $E^{1/3}$ and $E^{1/6}$ respectively, as expected. 

To summarise, we have seen that the solution close to an attractor without phase shift is well-described by our asymptotic solution  
 obtained by keeping only the logarithmic singularity 
contribution of the inviscid expression of the streamfunction.  This has been observed for all types of forcing, in 2D and in 3D, and for configurations where the attractor is connected to the critical latitude or not.

\begin{figure}
    \centering
    \includegraphics[width=\textwidth]{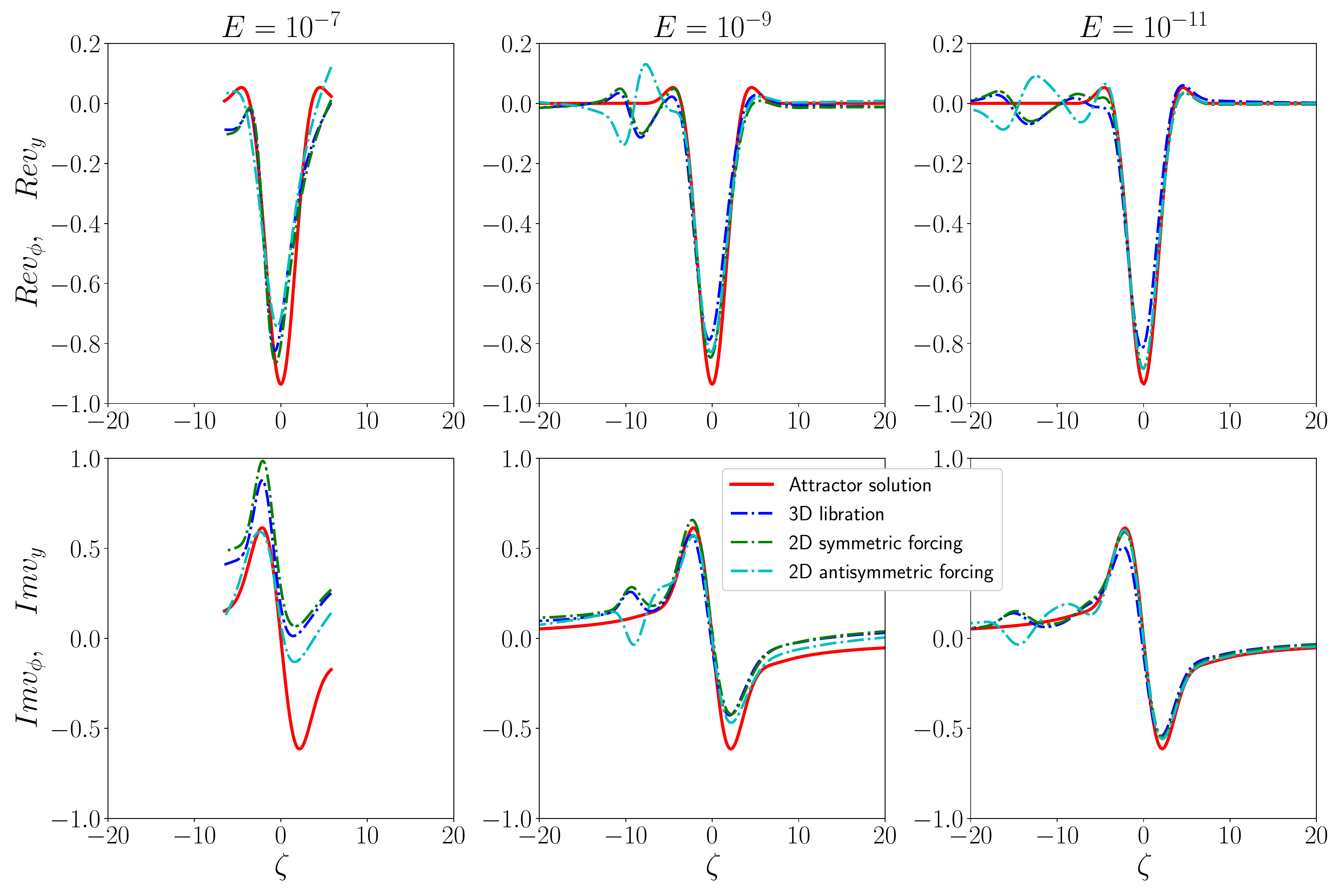}
    \caption{Comparison of velocity profiles between the attractor solutions and the numerical solutions on the cut $S_2$ of the equatorial attractor excited by the three forcings imposed on the outer core at three Ekman numbers.}
    \label{fig:eta=0.350_omega=0.8102_outercore_profile}
\end{figure}

\begin{figure}
    \centering
    \includegraphics[width=\textwidth]{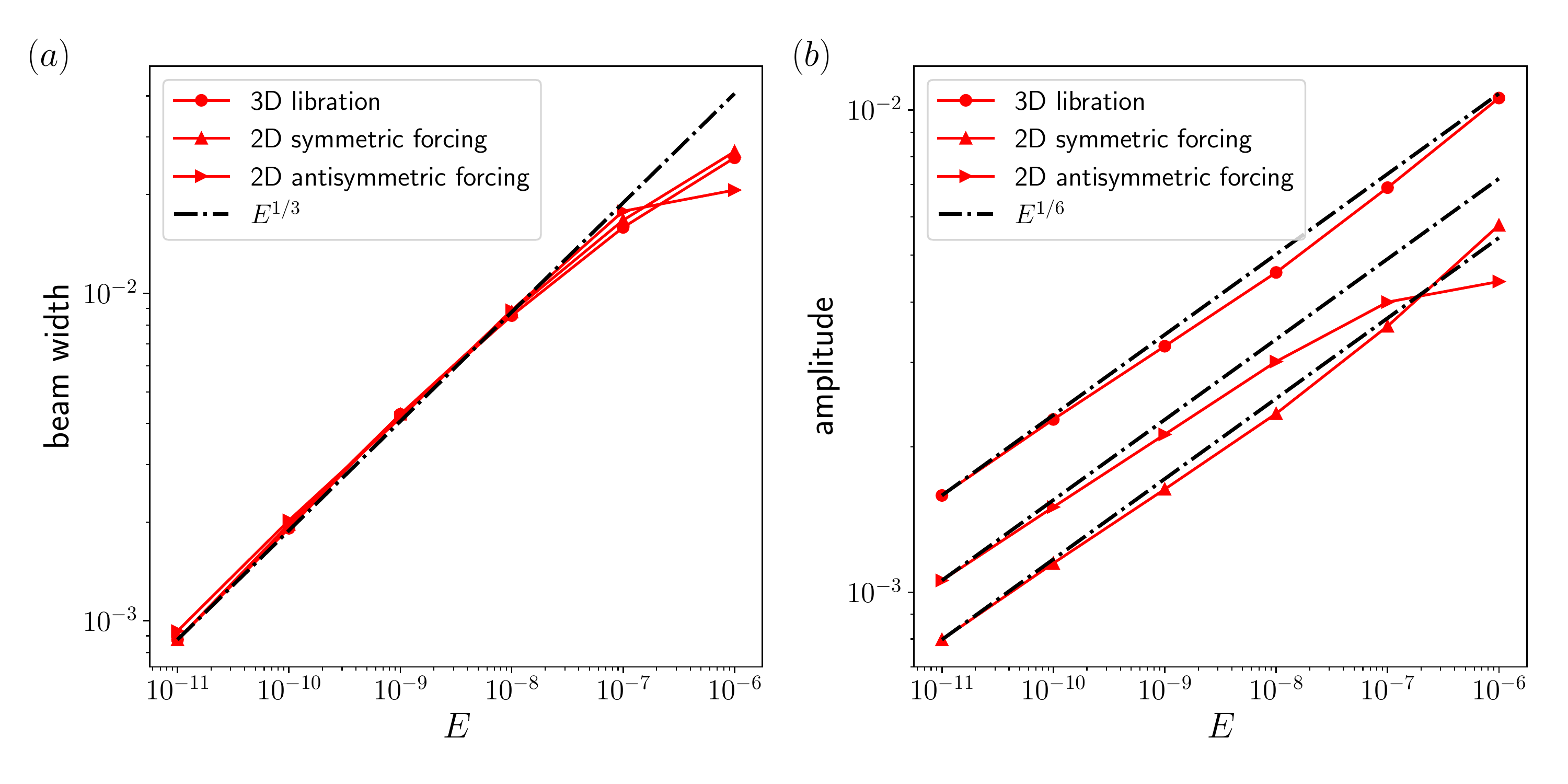}
    \caption{Ekman number scalings of the equatorial attractor excited by the three forcings imposed on the outer core: $(a)$ beam width measured by the distance of the peaks of the profiles in figure \ref{fig:eta=0.350_omega=0.8102_outercore_profile}; $(b)$ velocity amplitude taken at the critical position.}
    \label{fig:eta=0.350_omega=0.8102_outercore_scaling}
\end{figure}

\section{Conclusion}\label{sec:conclusion}
Using asymptotic analysis and numerical integration, we have studied the linear harmonic solution obtained in a rotating spherical shell  by librating the inner or outer core for very small Ekman numbers. We have 
considered a shell aspect ratio and a forcing frequency such that the ray beams converge towards either a polar attractor touching the vertical axis, or an equatorial attractor not touching the vertical axis.  
Both 3D axisymmetric and 2D configurations with different types of forcing have been
considered to analyse the effect of the  geometric singularity on the axis (obtained in 3D) and the influence of a phase shift  (present in the polar attractor in 3D and in 2D with a symmetric forcing). 

We have focused our interest on the concentrated internal shear layers that appear along the ray emitted from the critical latitude on the inner core, and close to the attractors.

We have first shown,  that, when the forcing is performed on the inner core, the dominant part of the solution is associated with a critical latitude beam. We have shown that the
characteristics of this beam are obtained by propagating the self-similar solution issued from the critical latitude on the inner core, as in an unbounded geometry \citep{ledizesInternalShearLayers2017} 
or for periodic ray paths (HFRL22).  This self-similar solution has a width in $E^{1/3}$, a well-defined  velocity amplitude in $E^{1/12}$, and a velocity structure corresponding to the
singularity index $m=5/4$.  As it propagates and reflects on boundaries (and possibly on the axis), its amplitude decreases down to $E^{1/6}$ until it reaches one of the two attractors. 

We have then observed that the numerical solution departs from the asymptotic critical-latitude solution when we get close to the attractor, for some of the attractors. 
We have seen that the departure was present when the rays do not exhibit a phase shift along the attractor, that is for the equatorial attractor and for the polar attractor in 2D with an antisymmetric forcing. 
We have then constructed a new asymptotic solution to describe the solution close to such an attractor, using results from O05. 
The main idea is based on the derivation of an inviscid functional equation for the streamfunction obtained by propagating the solution on a complete cycle on the attractor 
taking into account contraction/expansion as well as  Ekman pumping from the boundaries.  
The equation that is obtained when there is no phase shift is the equation obtained by O05. 
We have solved this equation by keeping only the logarithmic singular part. When smoothed by viscosity, this singular behavior leads to  a self-similar expression
for the velocity with a singularity index $m=1$ and an amplitude in $E^{1/6}$. 
Contrarily to the critical-latitude solution, the amplitude of this attractor solution depends on the Ekman pumping at the locations where the attractor touches the boundaries. 
 We have shown that it describes correctly the numerical solution close to the attractor for all the attractors without phase shift. 

From an asymptotic point of view,  it would now be useful to obtain a solution which describes both the critical solution and the attractor solution in order to  understand 
how the index characterising the self-similar solution changes from $m=5/4$ and $m=1$. 

When the attractor exhibits a phase shift, the analysis of O05 cannot be completely applied. 
We have seen that a different functional  equation is obtained for the streamfunction 
which does not possess any logarithmically singular solution. We suspect that the amplitude of the solution is weaker in that case which could explain
why its contribution is not visible in the numerical solution close to the attractor. 
Obtaining an asymptotic expression describing the solution in that case still constitutes one of the important remaining issues.

% \backsection[Supplementary data]{\label{SupMat}Supplementary material and movies are available at \\https://doi.org/10.1017/jfm.2019...}

\backsection[Acknowledgements]{J. He acknowledges China Scholarship Council for financial support (CSC 202008440260). Centre de Calcul Intensif d'Aix-Marseille is acknowledged for granting access to its high performance computing resources.}

% \backsection[Funding]{Please provide details of the sources of financial support for all authors, including grant numbers. For example, "This work was supported by the National Science Foundation (grant number XXXXXXX)". Multiple grant numbers should be separated by a comma and space, and where research was funded by more than one agency the different agencies should be separated by a semi-colon, with 'and' before the final funder. Grants held by different authors should be identified as belonging to individual authors by the authors' initials. For example, "This work was supported by the Deutsche Forschungsgemeinschaft (A.B., grant numbers XXXX, YYYY), (C.D., grant number ZZZZ); the Natural Environment Research Council (E.F., grant number FFFF); and the Australian Research Council (A.B., grant number GGGG), (E.F., grant number HHHH)". \\
% Where no specific funding has been provided for research, please provide the following statement: "This research received no specific grant from any funding agency, commercial or not-for-profit sectors." }

\backsection[Declaration of interests]{ The authors report no conflict of interest.}

% \backsection[Data availability statement]{The data that support the findings of this study are openly available in [repository name] at http://doi.org/[doi], reference number [reference number].}

% \backsection[Author ORCID]{Authors may include the ORCID identifers as follows.  F. Smith, https://orcid.org/0000-0001-2345-6789; B. Jones, https://orcid.org/0000-0009-8765-4321}

% \backsection[Author contributions]{Authors may include details of the contributions made by each author to the manuscript, for example, ``A.G. and T.F. derived the theory and T.F. and T.D. performed the simulations.  All authors contributed equally to analysing data and reaching conclusions, and in writing the paper.''}

\appendix

\section{Verification of the spectral codes}\label{app:verification}

\begin{figure}
    \centering
    \includegraphics[width=0.6\textwidth]{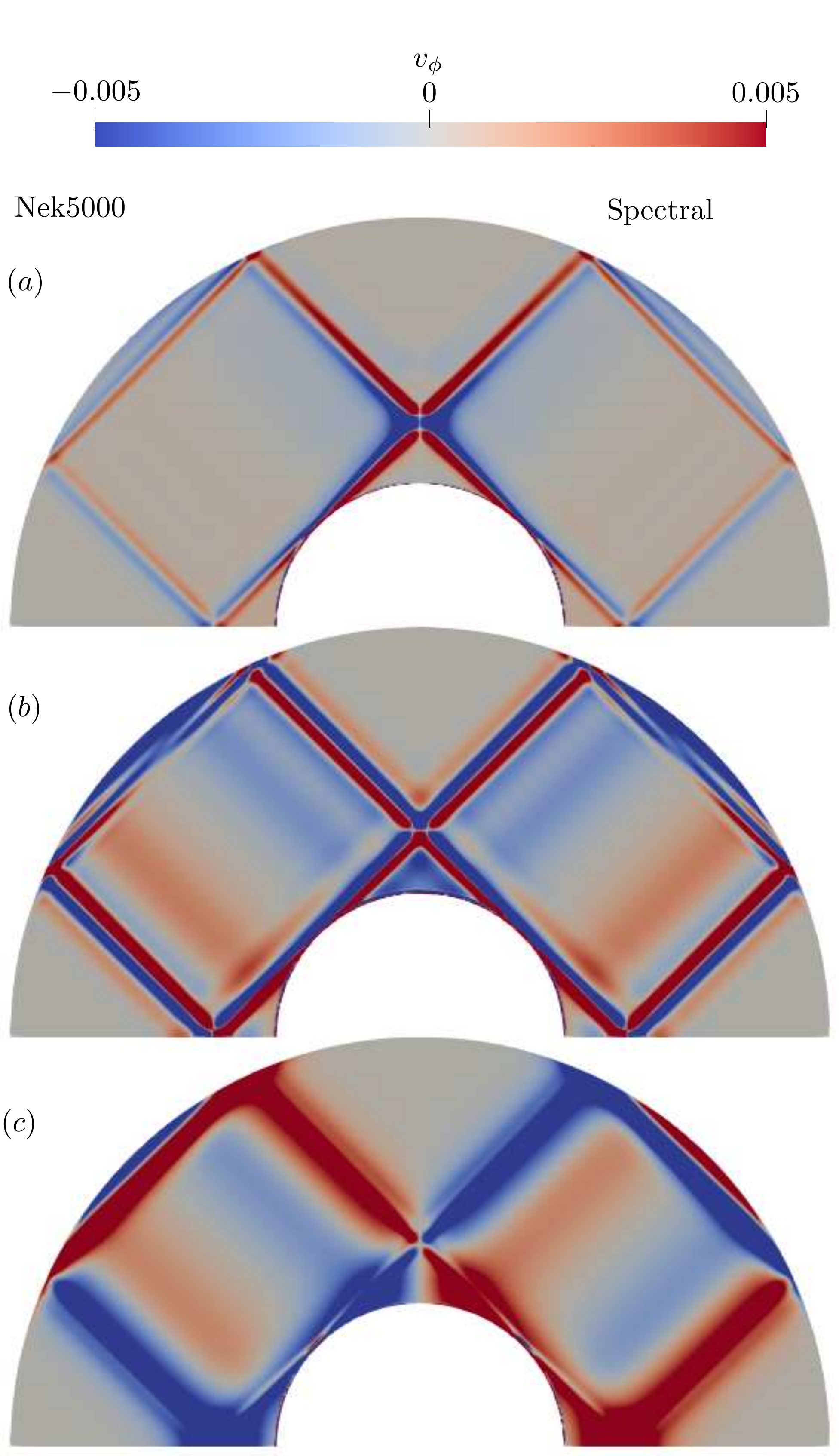}
    \caption{Comparison between the direct numerical results given by Nek5000 and the spectral codes: $(a)$ 3D libration; $(b)$ 2D symmetric forcing; $(c)$ 2D antisymmetric forcing. The combination of the aspect ratio and forcing frequency is $(\eta,\omega)=(0.35,\sqrt{2})$.}
    \label{fig:nek_spectral_contour}
\end{figure}

The spectral codes are verified against the open-source spectral-element software NEK5000 \citep{fischer1997overlapping} (NEK5000 Version 19.0, Argonne National Laboratory, Illinois;
available at https://nek5000.mcs.anl.gov). This code has already been used in the context of inertial wave propagation \citep{favierNonlinearEvolutionTidally2014}. Linear temporal simulations with the time-harmonic forcing are implemented by NEK5000. After enough number of periods, the time-harmonic steady state is reached and the results at different instants are extracted to compare with the real and the imaginary parts of the spectral results. Since it is almost impossible to reach the very low Ekman number $10^{-11}$ when solving the initial value problem with Nek5000, the comparison is done at the relative high Ekman number $10^{-6}$. The simulations are run for all the three forcings considered in this work. In the 3D configuration, the simulation is run in the upper-right quarter of an annulus, with the axisymmetric and symmetric boundary conditions set on the two straight boundaries respectively. In the 2D configuration, the simulations are run in the upper half of an annulus, with symmetric boundary conditions set on the two straight boundaries. One of the curved boundaries is subject to the harmonic forcing, while the other is subject to the no-slip boundary condition. The aspect ratio and the forcing frequency are chosen to be $0.35$ and $\sqrt{2}$ respectively, so that the wave pattern is a simple periodic orbit as in HFRL22. The comparisons are shown in figure \ref{fig:nek_spectral_contour}. The results of NEK5000 are shown on the left side, while those of the spectral codes are shown on the right side. They agree with each other very well.

\begin{figure}
    \centering
    \includegraphics[width=\textwidth]{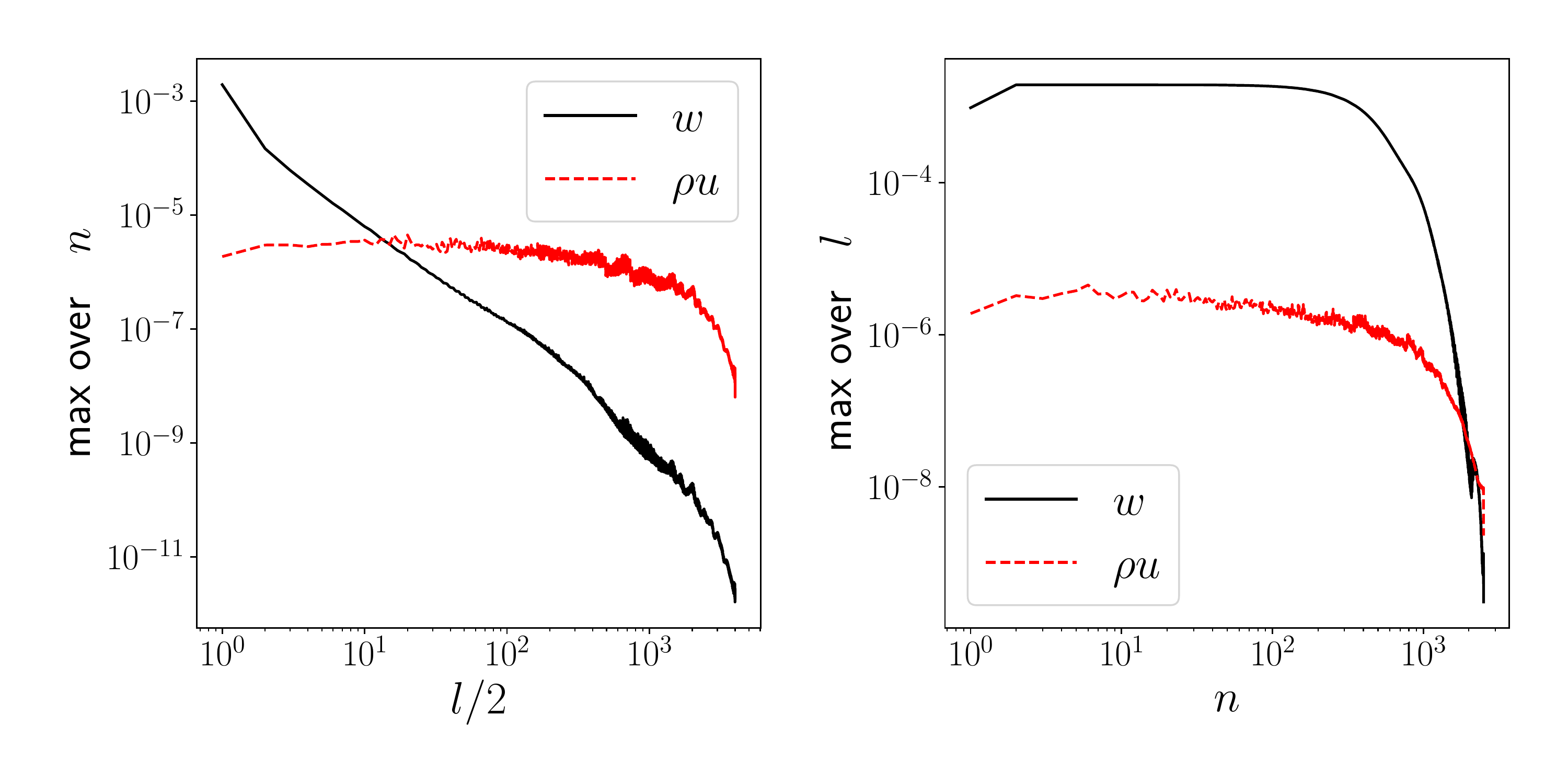}
    \caption{Spectra of the spherical harmonic components ($a$) and the Chebyshev coefficients ($b$) at $E=10^{-11}$ with the resolution $(N,L)=(2500,8000)$. For each $l$ or $n$, the maximum value over the other spectral component is taken. The forcing corresponds to the 3D libration imposed on the inner core. The aspect ratio is $\eta=0.35$ and the forcing frequency is $\omega=0.8102$.}
    \label{fig:spectra}
\end{figure}

On the other hand, the convergence of the spectral codes is verified by the spectra of the spherical harmonic (or Fourier) components and the Chebyshev coefficients, as in \citet{rieutordInertialWavesRotating1997}. 
Figure \ref{fig:spectra} shows the spectra for the 3D libration imposed on the inner core with the aspect ratio $0.35$ and forcing frequency $0.8102$ at the lowest Ekman number $10^{-11}$. 
The 2D results are similar and omitted.

\section{Ekman pumping}\label{app:Ekman_pumping}
The viscous forcing generates an Ekman layer adjacent to the boundary. The Ekman pumping plays a role in the generation of wave beams in the bulk. In order to derive the formula of it, it is convenient to use the streamfunction expression in the spherical or polar coordinates.
\subsection{3D configuration}
We first consider the libration imposed on the inner core. 
In the spherical coordinates $(\rho,\theta,\phi)$, the streamfunction $\psi$ and the associated variable $\chi$ are defined as
\begin{equation}\label{eq:3d_streamfunction_definition_spherical}
    v_\rho=\frac{1}{\rho^2\sin{\theta}}\frac{\partial\psi}{\partial\theta}, \quad v_\theta=-\frac{1}{\rho\sin{\theta}}\frac{\partial\psi}{\partial\rho}, \quad v_\phi=\frac{\chi}{\rho\sin{\theta}}.
\end{equation}
The governing equations (\ref{eq:governing_equations_vector_form}) is recast to
\begin{subeqnarray}\label{eq:3d_governing_equations_streamfunction_spherical}
    -\mathrm{i}\omega D^2\psi+2(\cos{\theta}\frac{\partial\chi}{\partial\rho}-\frac{\sin{\theta}}{\rho}\frac{\partial\chi}{\partial\theta})-ED^4\psi=0,\\[3pt]
    -\mathrm{i}\omega\chi-2(\cos{\theta}\frac{\partial\psi}{\partial \rho}-\frac{\sin{\theta}}{\rho}\frac{\partial\psi}{\partial\theta})-ED^2\chi=0,
\end{subeqnarray}
with the operator 
\begin{equation}
    D^2=\frac{\partial^2}{\partial \rho^2}-\frac{1}{\rho^2\tan\theta}\frac{\partial}{\partial\theta}+\frac{1}{\rho^2}\frac{\partial^2}{\partial\theta^2},
\end{equation}
and the boundary conditions
\begin{subeqnarray}
    \psi=\partial\psi/\partial\rho=0, \quad \chi=\eta^2\sin^2\theta \quad \mbox{at} \quad \rho=\eta, \\[3pt]
    \psi=\partial\psi/\partial\rho=\chi=0 \quad \mbox{at} \quad \rho=1.
\end{subeqnarray}
% The libration is considered to be imposed on the inner core first. 

The length scale of the Ekman layer is $\sqrt{E}$. The radial distance to the centre is rescaled as
\begin{equation}
    \hat{\rho}=(\rho-\eta)/\sqrt{E}.
\end{equation}
The streamfunction $\psi$ and the associated $\chi$ are expanded as to the leading order
\begin{equation}
    \psi=\sqrt{E}\hat{\psi}^{(1)}(\hat{\rho},\theta), \quad \chi=\hat{\chi}^{(0)}(\hat{\rho},\theta).
\end{equation}
In the leading order, the governing equations (\ref{eq:3d_governing_equations_streamfunction_spherical}) become
\begin{subeqnarray}
    -\mathrm{i}\omega\frac{\partial^2\hat{\psi}^{(1)}}{\partial\hat{\rho}^2}+2\cos{\theta}\frac{\partial\hat{\chi}^{(0)}}{\partial\hat{\rho}}-\frac{\partial^4\hat{\psi}^{(1)}}{\partial\hat{\rho}^4}=0,\\
    -\mathrm{i}\omega\hat{\chi}^{(0)}-2\cos{\theta}\frac{\partial\hat{\psi}^{(1)}}{\partial\hat{\rho}}-\frac{\partial^2\hat{\chi}^{(0)}}{\partial\hat{\rho}^2}=0,
\end{subeqnarray}
with the boundary conditions
\begin{subeqnarray}
    \psi=\partial\psi/\partial\rho=0, \quad \chi=\eta^2\sin^2\theta, \quad \mbox{as} \quad \hat{\rho}=0,\\[3pt]
    \partial\psi/\partial\rho\rightarrow0, \quad \chi\rightarrow0, \quad \mbox{as} \quad \hat{\rho}\rightarrow \infty.
\end{subeqnarray}
The solution of the streamfunction is obtained as
\begin{equation}
    \psi=\sqrt{E}\mathrm{i}\frac{\eta^2\sin^2\theta}{2}\left(-\frac{e^{-\lambda_+\hat{\rho}}}{\lambda_+}+\frac{1}{\lambda_+}+\frac{e^{-\lambda_-\hat{\rho}}}{\lambda_-}-\frac{1}{\lambda_-}\right),
\end{equation}
with $\lambda_\pm$ defined as
\begin{equation}
    \lambda_+=(1-i)\sqrt{\omega/2+\cos{\theta}}
\end{equation}
and
\begin{equation}
 \lambda_- = \left\{
    \begin{array}{ll}
      (1-i)\sqrt{\omega/2-\cos{\theta}}, & \omega/2>\cos{\theta}, \\[2pt]
      (1+i)\sqrt{\cos{\theta}-\omega/2}, & \omega/2<\cos{\theta}.
  \end{array} \right.
\end{equation}
When $\hat{\rho}$ goes to $+\infty$, the Ekman pumping is obtained as
\begin{equation}
    \psi^{(EP)}=\frac{\mathrm{i}\eta^2\sin^2\theta}{2}\left(\frac{1}{\lambda_+}-\frac{1}{\lambda_-}\right)\sqrt{E}.
\end{equation}
The Ekman pumping blows up at the critical co-latitude $\theta_c=\arccos{\omega/2}$.

When the libration is imposed on the outer core, the boundary conditions become different and the corresponding Ekman pumping can be obtained similarly, which is
\begin{equation}
    \psi^{(EP)}=\frac{-\mathrm{i}\sin^2\theta}{2}\left(\frac{1}{\lambda_+}-\frac{1}{\lambda_-}\right)\sqrt{E}.
\end{equation}

\subsection{2D configuration}
In the 2D configuration, the governing equations (\ref{eq:2d_governing_equations_streamfunction_polar}) of the streamfunction and the associated variable $\chi$  in the polar coordinates are solved asymptotically using the same scaling $E^{1/2}$ as in the 3D configuration.
The expressions of the Ekman pumping generated by different forcings are given in table \ref{tab:Ekman_pumping}. Note that $\vartheta$ in the 2D configuration has been replaced by $\pi/2-\theta$ in order to keep the similar expressions as in the 3D counterpart.

\begin{table}
  \begin{center}
\def~{\hphantom{0}}
  \begin{tabular}{lccc}
                      & 3D libration  & 2D symmetric forcing   &   2D antisymmetric forcing \\[3pt]
       inner core & $\frac{\mathrm{i}\eta^2\sin^2\theta}{2}\left(\frac{1}{\lambda_+}-\frac{1}{\lambda_-}\right)\sqrt{E}$  & $\frac{\mathrm{i}}{2}\left(\frac{1}{\lambda_+}-\frac{1}{\lambda_-}\right)\sqrt{E}$ & $\frac{\mathrm{i}\eta\sin\theta}{2}\left(\frac{1}{\lambda_+}-\frac{1}{\lambda_-}\right)\sqrt{E}$ \\
       outer core & $\frac{-\mathrm{i}\sin^2\theta}{2}\left(\frac{1}{\lambda_+}-\frac{1}{\lambda_-}\right)\sqrt{E}$   & $\frac{-\mathrm{i}}{2}\left(\frac{1}{\lambda_+}-\frac{1}{\lambda_-}\right)\sqrt{E}$ & $\frac{-\mathrm{i}\sin\theta}{2}\left(\frac{1}{\lambda_+}-\frac{1}{\lambda_-}\right)\sqrt{E}$ \\
  \end{tabular}
  \caption{Expressions of Ekman pumping $\psi^{(EP)}$ generated by different forcings.}
  \label{tab:Ekman_pumping}
  \end{center}
\end{table}

Note that, the Ekman pumping is $O(E^{1/2})$, except at the critical latitude where the Ekman pumping blows up.

\bibliographystyle{jfm}
\bibliography{jfm}

\begin{thebibliography}{38}
\expandafter\ifx\csname natexlab\endcsname\relax\def\natexlab#1{#1}\fi
\def\au#1{#1} \def\ed#1{#1} \def\yr#1{#1}\def\at#1{#1}\def\jt#1{\textit{#1}}
  \def\bt#1{#1}\def\bvol#1{\textbf{#1}} \def\vol#1{#1} \def\pg#1{#1}
  \def\publ#1{#1}\def\arxiv#1{#1}\def\org#1{#1}\def\st#1{\textit{#1}}

\bibitem[Bezanson {\em et~al.\/}(2017)Bezanson, Edelman, Karpinski \&
  Shah]{bezanson2017julia}
{\sc \au{Bezanson, J.}, \au{Edelman, A.}, \au{Karpinski, S.} \& \au{Shah,
  V.~B.}} \yr{2017}  \at{Julia: A fresh approach to numerical computing}.
  \jt{SIAM Rev.}  \bvol{59}~(1),  \pg{65--98}.

\bibitem[Engeln-M{\`e}ullges \&
  Uhlig(1996)]{engeln-meullgesNumericalAlgorithms1996}
{\sc \au{Engeln-M{\`e}ullges, G.} \& \au{Uhlig, F.}} \yr{1996} {\em Numerical
  algorithms with C\/}.  \publ{Springer Science \& Business Media}.

\bibitem[Favier {\em et~al.\/}(2014)Favier, Barker, Baruteau \&
  Ogilvie]{favierNonlinearEvolutionTidally2014}
{\sc \au{Favier, B.}, \au{Barker, A.~J.}, \au{Baruteau, C.} \& \au{Ogilvie,
  G.~I.}} \yr{2014}  \at{Non-linear evolution of tidally forced inertial waves
  in rotating fluid bodies}.  \jt{Mon. Not. R. Astron. Soc.}  \bvol{439}~(1),
  \pg{845--860}.

\bibitem[Fischer(1997)]{fischer1997overlapping}
{\sc \au{Fischer, P.~F.}} \yr{1997}  \at{An overlapping schwarz method for
  spectral element solution of the incompressible navier--stokes equations}.
  \jt{J. Comput. Phys.}  \bvol{133}~(1),  \pg{84--101}.

\bibitem[Flynn {\em et~al.\/}(2003)Flynn, Onu \&
  Sutherland]{flynnInternalWaveExcitation2003}
{\sc \au{Flynn, M.~R.}, \au{Onu, K.} \& \au{Sutherland, B.~R.}} \yr{2003}
  \at{Internal wave excitation by a vertically oscillating sphere}.  \jt{J.
  Fluid Mech.}  \bvol{494},  \pg{65--93}.

\bibitem[Ghaemsaidi \& Peacock(2013)]{ghaemsaidi3DStereoscopicPIV2013}
{\sc \au{Ghaemsaidi, S.~J.} \& \au{Peacock, T.}} \yr{2013}  \at{{{3D
  Stereoscopic PIV}} visualization of the axisymmetric conical internal wave
  field generated by an oscillating sphere}.  \jt{Exp. Fluids}  \bvol{54}~(2),
  \pg{1454}.

\bibitem[Greenspan(1968)]{greenspan1968theory}
{\sc \au{Greenspan, H.~P.}} \yr{1968} {\em The theory of rotating fluids\/}.
  \publ{CUP Archive}.

\bibitem[He {\em et~al.\/}(2022)He, Favier, Rieutord \&
  Le~Diz{\`e}s]{heInternalShearLayers2022}
{\sc \au{He, J.}, \au{Favier, B.}, \au{Rieutord, M.} \& \au{Le~Diz{\`e}s, S.}}
  \yr{2022}  \at{Internal shear layers in librating spherical shells: The case
  of periodic characteristic paths}.  \jt{J. Fluid Mech.}  \bvol{939}.

\bibitem[Hurley(1997)]{hurleyGenerationInternalWaves1997a}
{\sc \au{Hurley, D.~G.}} \yr{1997}  \at{The generation of internal waves by
  vibrating elliptic cylinders. {{Part}} 1. {{Inviscid}} solution}.  \jt{J.
  Fluid Mech.}  \bvol{351},  \pg{105--118}.

\bibitem[Hurley \& Keady(1997)]{hurleyGenerationInternalWaves1997}
{\sc \au{Hurley, D.~G.} \& \au{Keady, G.}} \yr{1997}  \at{The generation of
  internal waves by vibrating elliptic cylinders. {{Part}} 2. {{Approximate}}
  viscous solution}.  \jt{J. Fluid Mech.}  \bvol{351},  \pg{119--138}.

\bibitem[Jouve \& Ogilvie(2014)]{jouveDirectNumericalSimulations2014}
{\sc \au{Jouve, L.} \& \au{Ogilvie, G.~I.}} \yr{2014}  \at{Direct numerical
  simulations of an inertial wave attractor in linear and nonlinear regimes}.
  \jt{J. Fluid Mech.}  \bvol{745},  \pg{223--250}.

\bibitem[Kerswell(1995)]{kerswellInternalShearLayers1995}
{\sc \au{Kerswell, R.~R.}} \yr{1995}  \at{On the internal shear layers spawned
  by the critical regions in oscillatory {{Ekman}} boundary layers}.  \jt{J.
  Fluid Mech.}  \bvol{298},  \pg{311--325}.

\bibitem[Kida(2011)]{kidaSteadyFlowRapidly2011}
{\sc \au{Kida, S.}} \yr{2011}  \at{Steady flow in a rapidly rotating sphere
  with weak precession}.  \jt{J. Fluid Mech.}  \bvol{680},  \pg{150--193}.

\bibitem[Le~Diz{\`e}s(2020)]{ledizesReflectionOscillatingInternal2020}
{\sc \au{Le~Diz{\`e}s, S.}} \yr{2020}  \at{Reflection of oscillating internal
  shear layers: Nonlinear corrections}.  \jt{J. Fluid Mech.}  \bvol{899},
  \pg{A21}.

\bibitem[Le~Diz{\`e}s(2022)]{LeDizes2023}
{\sc \au{Le~Diz{\`e}s, S.}} \yr{2022} Singularities from critical slopes.
  \bt{In {\em Singularities and attractors in rotating and stratified
  fluids\/}}. IRPHE, Marseille.

\bibitem[Le~Diz{\`e}s \& Le~Bars(2017)]{ledizesInternalShearLayers2017}
{\sc \au{Le~Diz{\`e}s, S.} \& \au{Le~Bars, M.}} \yr{2017}  \at{Internal shear
  layers from librating objects}.  \jt{J. Fluid Mech.}  \bvol{826},
  \pg{653--675}.

\bibitem[Lin \& Noir(2020)]{linLibrationdrivenInertialWaves2020}
{\sc \au{Lin, Y.} \& \au{Noir, J.}} \yr{2020}  \at{Libration-driven inertial
  waves and mean zonal flows in spherical shells}.  \jt{Geophys. Astrophys.
  Fluid Dyn.}  \bvol{0}~(0),  \pg{1--22}.

\bibitem[Maas {\em et~al.\/}(1997)Maas, Benielli, Sommeria \&
  Lam]{maasObservationInternalWave1997}
{\sc \au{Maas, L.}, \au{Benielli, D.}, \au{Sommeria, J.} \& \au{Lam, F.-P.}}
  \yr{1997}  \at{Observation of an internal wave attractor in a confined,
  stably stratified fluid}.  \jt{Nature}  \bvol{388}~(6642),  \pg{557--561}.

\bibitem[Maas \& Lam(1995)]{maasGeometricFocusingInternal1995}
{\sc \au{Maas, L. R.~M.} \& \au{Lam, F.~A.}} \yr{1995}  \at{Geometric focusing
  of internal waves}.  \jt{J. Fluid Mech.}  \bvol{300},  \pg{1--41}.

\bibitem[Machicoane {\em et~al.\/}(2015)Machicoane, Cortet, Voisin \&
  Moisy]{machicoaneInfluenceMultipoleOrder2015}
{\sc \au{Machicoane, N.}, \au{Cortet, P.}, \au{Voisin, B.} \& \au{Moisy, F.}}
  \yr{2015}  \at{Influence of the multipole order of the source on the decay of
  an inertial wave beam in a rotating fluid}.  \jt{Phys. Fluids}
  \bvol{27}~(6),  \pg{066602}.

\bibitem[Moore \& Saffman(1969)]{mooreStructureFreeVertical1969}
{\sc \au{Moore, D.~W.} \& \au{Saffman, P.~G.}} \yr{1969}  \at{The structure of
  free vertical shear layers in a rotating fluid and the motion produced by a
  slowly rising body}.  \jt{Philos. Trans. R. Soc. London, Ser. A}
  \bvol{264}~(1156),  \pg{597--634}.

\bibitem[Ogilvie(2005)]{ogilvieWaveAttractorsAsymptotic2005}
{\sc \au{Ogilvie, G.~I.}} \yr{2005}  \at{Wave attractors and the asymptotic
  dissipation rate of tidal disturbances}.  \jt{J. Fluid Mech.}  \bvol{543},
  \pg{19--44}.

\bibitem[Ogilvie(2009)]{ogilvieTidalDissipationRotating2009}
{\sc \au{Ogilvie, G.~I.}} \yr{2009}  \at{Tidal dissipation in rotating fluid
  bodies: A simplified model}.  \jt{Mon. Not. R. Astron. Soc.}  \bvol{396}~(2),
   \pg{794--806}.

\bibitem[Ogilvie(2020)]{ogilvie2020internal}
{\sc \au{Ogilvie, G.~I.}} \yr{2020}  \at{Internal waves and tides in stars and
  giant planets}.  \bt{In {\em Fluid Mechanics of Planets and Stars\/}},
  \pg{pp. 1--30}.  \publ{Springer}.

\bibitem[Ogilvie \& Lin(2004)]{ogilvieTidalDissipationRotating2004}
{\sc \au{Ogilvie, G.~I.} \& \au{Lin, D. N.~C.}} \yr{2004}  \at{Tidal
  {{Dissipation}} in {{Rotating Giant Planets}}}.  \jt{ApJ}  \bvol{610}~(1),
  \pg{477--509}.

\bibitem[Rieutord(1991)]{rieutord91}
{\sc \au{Rieutord, M.}} \yr{1991}  \at{{Linear theory of rotating fluids using
  spherical harmonics. II. Time periodic flows}}.  \jt{Geophys. Astrophys.
  Fluid Dyn.}  \bvol{59},  \pg{185--208}.

\bibitem[Rieutord {\em et~al.\/}(2000)Rieutord, Georgeot \&
  Valdettaro]{Rieutord2000}
{\sc \au{Rieutord, M.}, \au{Georgeot, B.} \& \au{Valdettaro, L.}} \yr{2000}
  \at{Wave {{Attractors}} in {{Rotating Fluids}}: {{A Paradigm}} for
  {{Ill-Posed Cauchy Problems}}}.  \jt{Phys. Rev. Lett.}  \bvol{85}~(20),
  \pg{4277--4280}.

\bibitem[Rieutord {\em et~al.\/}(2001)Rieutord, Georgeot \&
  Valdettaro]{rieutordInertialWavesRotating2001}
{\sc \au{Rieutord, M.}, \au{Georgeot, B.} \& \au{Valdettaro, L.}} \yr{2001}
  \at{Inertial waves in a rotating spherical shell: Attractors and asymptotic
  spectrum}.  \jt{J. Fluid Mech.}  \bvol{435},  \pg{103--144}.

\bibitem[Rieutord \& Valdettaro(1997)]{rieutordInertialWavesRotating1997}
{\sc \au{Rieutord, M.} \& \au{Valdettaro, L.}} \yr{1997}  \at{Inertial waves in
  a rotating spherical shell}.  \jt{J. Fluid Mech.}  \bvol{341},  \pg{77--99}.

\bibitem[Rieutord \& Valdettaro(2010)]{rieutordViscousDissipationTidally2010}
{\sc \au{Rieutord, M.} \& \au{Valdettaro, L.}} \yr{2010}  \at{Viscous
  dissipation by tidally forced inertial modes in a rotating spherical shell}.
  \jt{J. Fluid Mech.}  \bvol{643},  \pg{363--394}.

\bibitem[Rieutord {\em et~al.\/}(2002)Rieutord, Valdettaro \&
  Georgeot]{rieutordAnalysisSingularInertial2002}
{\sc \au{Rieutord, M.}, \au{Valdettaro, L.} \& \au{Georgeot, B.}} \yr{2002}
  \at{Analysis of singular inertial modes in a spherical shell: The slender
  toroidal shell model}.  \jt{J. Fluid Mech.}  \bvol{463},  \pg{345--360}.

\bibitem[Roberts \& Stewartson(1963)]{robertsSTABILITYMACLAURINSPHEROID1963}
{\sc \au{Roberts, P.~H.} \& \au{Stewartson, K.}} \yr{1963}  \at{{{On the
  stability of a {MacLaurin} spheroid of small viscosity}}}.  \jt{Astrophys.
  J.}  \bvol{137}~(3),  \pg{777--790}.

\bibitem[Sutherland \& Linden(2002)]{sutherlandInternalWaveExcitation2002}
{\sc \au{Sutherland, B.~R.} \& \au{Linden, P.~F.}} \yr{2002}  \at{Internal wave
  excitation by a vertically oscillating elliptical cylinder}.  \jt{Phys.
  Fluids}  \bvol{14}~(2),  \pg{721--731}.

\bibitem[Thomas \& Stevenson(1972)]{thomasSimilaritySolutionViscous1972}
{\sc \au{Thomas, N.~H.} \& \au{Stevenson, T.~N.}} \yr{1972}  \at{A similarity
  solution for viscous internal waves}.  \jt{J. Fluid Mech.}  \bvol{54}~(3),
  \pg{495--506}.

\bibitem[Tilgner(1999)]{tilgnerDrivenInertialOscillations1999}
{\sc \au{Tilgner, A.}} \yr{1999}  \at{Driven inertial oscillations in spherical
  shells}.  \jt{Phys. Rev. E}  \bvol{59}.

\bibitem[Voisin(2003)]{voisinLimitStatesInternal2003}
{\sc \au{Voisin, B.}} \yr{2003}  \at{Limit states of internal wave beams}.
  \jt{J. Fluid Mech.}  \bvol{496},  \pg{243--293}.

\bibitem[Voisin {\em et~al.\/}(2011)Voisin, Ermanyuk \&
  Fl{\'o}r]{voisinInternalWaveGeneration2011}
{\sc \au{Voisin, B.}, \au{Ermanyuk, E.~V.} \& \au{Fl{\'o}r, J.-B.}} \yr{2011}
  \at{Internal wave generation by oscillation of a sphere, with application to
  internal tides}.  \jt{J. Fluid Mech.}  \bvol{666},  \pg{308--357}.

\bibitem[Zhang {\em et~al.\/}(2007)Zhang, King \&
  Swinney]{zhangExperimentalStudyInternal2007}
{\sc \au{Zhang, H.~P.}, \au{King, B.} \& \au{Swinney, Harry~L.}} \yr{2007}
  \at{Experimental study of internal gravity waves generated by supercritical
  topography}.  \jt{Phys. Fluids}  \bvol{19}~(9),  \pg{096602}.

\end{thebibliography}
%Use of the above commands will create a bibliography using the .bib file. Shown below is a bibliography built from individual items.

% \bibliographystyle{jfm}
%\bibliography{jfm2esam}

%% End of file `jfm2esam.bib'.

\end{document}